\documentclass[amsmath,amssymb]{revtex4}
\usepackage{amssymb,amsmath,graphicx}
\def\[{\begin{equation}}
\def\]{\end{equation}}
\def\x{\textbf{x}}
\def\u{\textbf{u}}
\def\F{\textbf{F}}
\def\L{\textbf{L}}
\def\M{\textbf{M}}
\def\P{\textbf{P}}
\def\G{\textbf{G}}
\def\rd{{\rm d}}
\advance\textheight -16mm
\begin{document}
\title{A numerical method for computing time-periodic solutions in dissipative wave systems}
\author{Jianke Yang}
\address{Department of Mathematics and Statistics, University of Vermont, Burlington, VT 05401, USA
\\ email: jyang@math.uvm.edu}

\begin{abstract}
A numerical method is proposed for computing time-periodic and
relative time-periodic solutions in dissipative wave systems. In
such solutions, the temporal period, and possibly other additional
internal parameters such as the propagation constant, are unknown
priori and need to be determined along with the solution itself. The
main idea of the method is to first express those unknown parameters
in terms of the solution through quasi-Rayleigh quotients, so that
the resulting integro-differential equation is for the time-periodic
solution only. Then this equation is computed in the combined
spatiotemporal domain as a boundary value problem by
Newton-conjugate-gradient iterations. The proposed method applies to
both stable and unstable time-periodic solutions; its numerical
accuracy is spectral; it is fast-converging; and its coding is short
and simple. As numerical examples, this method is applied to the
Kuramoto-Sivashinsky equation and the cubic-quintic Ginzburg-Landau
equation, whose time-periodic or relative time-periodic solutions
with spatially-periodic or spatially-localized profiles are
computed. This method also applies to systems of ordinary
differential equations, as is illustrated by its simple computation
of periodic orbits in the Lorenz equations. MATLAB codes for all
numerical examples are provided in appendices to illustrate the
simple implementation of the proposed method.
\end{abstract}

%Keywords: time-periodic solutions, dissipative wave systems,
%iteration methods


\maketitle

\section{Introduction}

In studies of nonlinear waves in physical systems, coherent
structures play a prominent role. The simplest coherent structures
are stationary or traveling waves, which do not change their shape
upon propagation. A familiar example is solitary waves in various
physical wave equations (such as the Korteweg-de Vries equation).
Another important class of coherent structures is time-periodic
solutions, which change their shape periodically upon propagation. A
familiar example is breathers in the sine-Gordon equation. Coherent
structures are important for nonlinear wave equations for obvious
reasons. If these structures are stable, they would serve as
attractors and dictate solution dynamics. Even if they are unstable,
they could still exert strong influence on the dynamical outcome
(such as contributing to chaotic behaviors). Thus determination of
coherent structures is a fundamental step toward the understanding
of nonlinear wave systems. This determination is often numerical due
to lack of analytical expressions.

If the wave system is conservative (i.e., without gain or loss),
these coherent structures generally exist as continuous families,
parameterized by their energy (or a related parameter such as wave
height). Solitary waves in the Korteweg-de Vries equation are such
examples, where the height of the wave is a free parameter. If the
wave system is dissipative, however, these coherent structures
generally exist as isolated objects, at discrete energy levels, due
to the requirement that the gain and loss of the energy must balance
each other exactly. Solitary waves in the complex Ginzburg-Landau
equation are such examples, where the height of the solitary wave is
fixed \cite{Akhmediev2001}.

Numerical computations of stationary and traveling waves in
nonlinear systems has a long history, and a large number of
effective numerical methods have been developed (see \cite{Yang2010}
and the references therein). Most of these methods were designed for
conservative systems, but some methods for dissipative systems are
also available \cite{Akhmediev2001,YangSOM}.

In this article, we consider computations of time-periodic solutions
in dissipative wave systems. This computation is more challenging
than in conservative systems, because the solution's temporal
period, as well as possibly other additional parameters, is
discrete, but such parameters are not known priori and have to be
computed together with the solution itself. So far, several
numerical methods have been used for these computations. If the
solution is stable, then it can be obtained as the long-time limit
of an initial value problem by evolution simulation. This evolution
method is often slow. More seriously, it cannot access unstable
solutions, which are needed in many situations (such as a
bifurcation study or estimation of fractal dimensions of a chaotic
attractor \cite{Artuso1990}). A second method is the damped Newton's
method, which was used on the Kuramoto-Sivashinsky equation
\cite{Zoldi1998}. In this method, the Kuramoto-Sivashinsky equation
was discretized in the spatiotemporal domain by finite differences,
and the resulting system of algebraic equations was solved by damped
Newton iterations. A third method is based on error minimization and
infinitesimal damped Newton iterations \cite{Cvitanovic2004}. This
method was developed for ordinary differential equations (ODEs), and
then applied to the Kuramoto-Sivashinsky equation after it was
converted to a system of ODEs through Fourier-series expansion. A
fourth method was used for computing relative time-periodic
solutions of the complex Ginzburg-Landau equation under periodic
boundary conditions \cite{Lopez2005}. In this method, the solution
was expanded into space-time Fourier series, so that the
Ginzburg-Landau equation was converted into a system of nonlinear
algebraic equations, which was then solved by a nonlinear least
squares solver from the MINPACK software package. One more method
was used for computing time-periodic and space-localized solutions
in the damped-driven nonlinear Schr\"odinger equation
\cite{Barashenkov2011}. In this method, the wave equation was
dicretized in the spatiotemporal domain by finite differences into a
set of nonlinear algebraic equations, which was then solved by
Newton iterations.

Computations of time-periodic solutions in partial differential
equations (PDEs) is closely related to computations of periodic
orbits in systems of ODEs. For systems of ODEs, quite a few
numerical methods are available. Examples include the multipoint
shooting method \cite{Shooting}, the finite-difference
discretization method, the collocation method
\cite{Ascher1979,LingWu1987}, the multipoint-shooting with automatic
differentiation method \cite{Gucken2000}, and so on (the software
package AUTO uses the B-spline collocation method \cite{AUTO}). In
principle, all these ODE-based methods can be adapted to PDEs if the
PDEs are first converted into a system of ODEs (by finite difference
or spatial-mode expansion). However, the extra cost of PDE-to-ODE
conversion and the inevitable large size of the resulting ODE system
make such methods not ideal for PDE applications.

In this article, we develop a new numerical method for computing
time-periodic and relative time-periodic solutions in dissipative
wave equations. This work is motivated by several reasons. First,
our view is that the best way to compute such solutions in PDEs is
to do so in the PDE framework, rather than converting PDEs to large
systems of ODEs or algebraic equations. The advantage of the PDE
framework is that the structure of the PDE is retained, and
important quantities such as the linearization operator of the PDE
can be calculated analytically. Second, almost all numerical methods
for time-periodic solutions in PDEs involve solving large systems of
linear equations. Since conjugate-gradient methods are widely
recognized as probably the fastest numerical method for solving
linear algebraic and operator equations \cite{Golub}, we are
motivated to incorporate conjugate-gradient methods into our
algorithm. Thirdly, a good numerical algorithm should also be simple
to implement. Since none of the previous numerical schemes provided
sample codes for the readers to peruse, we are motivated to provide
a set of simple sample codes, so that the readers can directly use
them, or modify them for their own problems. Building upon our
previous experience in designing Newton-conjugate-gradient methods
for computing solitary waves and their linear-stability eigenvalues
\cite{YangCG,Yang2010}, we now develop a method for computing
time-periodic solutions which meet the above goals.

The main idea of this proposed method is the following. In view of
the fact that the temporal period and possibly other additional
parameters in the time-periodic solutions are unknown priori, our
first step is to express those unknown parameters in terms of the
solution through quasi-Rayleigh quotients, so that the resulting
integro-differential equation is for the time-periodic solution
only. Then this equation is solved in the combined spatiotemporal
domain as a boundary value problem by the Newton-conjugate-gradient
method, where Newton corrections are obtained by preconditioned
conjugate-gradient iterations. The benefit of using
conjugate-gradient iterations to solve the Newton-correction
equation is two-fold: one is that it allows the computation to be
performed entirely in the PDE framework (since these iterations
apply to linear operator equations as well as matrix equations); and
the other is that the power of conjugate-gradient iterations for
solving large systems of linear equations can be brought out.

The proposed method applies to both stable and unstable
time-periodic solutions; its numerical accuracy is spectral (since
it is compatible with spectral differentiation
\cite{Trefethen2000,Boyd2001}), and its coding is short and simple.
As numerical examples, this method is applied to the
Kuramoto-Sivashinsky equation and the cubic-quintic Ginzburg-Landau
equation, whose time-periodic or relative time-periodic solutions
with spatially-periodic or spatially-localized profiles are
computed. This method also applies to systems of ODEs, as is
illustrated by its simple computation of periodic orbits in the
Lorenz equations. These numerical examples reveal that the proposed
method is very fast, as it only takes from a fraction of a second to
a couple of minutes (on a personal computer) to find solutions of
varying complexities to the accuracy of $10^{-10}$. The simplicity
of coding of the proposed method is evidenced in appendices, where
stand-alone MATLAB codes for all numerical examples are provided.

\section{A numerical method for time-periodic solutions with an unknown
period only}

We first present a numerical method for computing time-periodic
solutions whose temporal period is the only unknown parameter. For
example, time-periodic solutions in the Kuramoto-Sivashinsky
equation \cite{Kuramoto1976,Sivashinsky1977}
\[  \label{e:KS}
u_t+uu_x+u_{xx}+\gamma u_{xxxx}=0
\]
and the damped parametrically-driven nonlinear Schr\"odinger (NLS)
equation \cite{Barashenkov2011}
\[  \label{e:DDNLS}
i\psi_t+\psi_{xx}+2|\psi|^2\psi-\psi=h\psi^*-i\gamma \psi
\]
belong to this category. In the Kuramoto-Sivashinsky equation
(\ref{e:KS}), $u(x,t)$ is a real variable and $\gamma$ a real
``superviscosity" coefficient. In the damped-forced NLS equation
(\ref{e:DDNLS}), $\psi(x,t)$ is a complex variable, and $h, \gamma$
are real coefficients. Both equations admit solutions that are
time-periodic, but the temporal period is not known priori and needs
to be determined along with the solution itself
\cite{Barashenkov2011, Cvitanovic2004, Hyman1986, Kevrekidis1990,
Zoldi1998}.

Dissipative systems which admit time-periodic solutions with only an
unknown temporal period can be cast in the following general form
\[ \label{e:ut1}
\u_t=\F(\x, \partial_{\x}, \u),
\]
where $\x=(x_1, x_2, \cdots, x_N)$ is the $N$-dimensional spatial
coordinate, $\u(\x, t)$ is a real-valued vector variable of $\x$ and
time $t$, and $\F$ is a real-valued, generally nonlinear vector
function of $\x$, $\u$ and its spatial derivatives. Notice here that
we allow $\F$ to contain explicit dependence on $\x$ (to incorporate
spatial inhomogeneities), but not on time $t$. The
Kuramoto-Sivashinsky equation (\ref{e:KS}) naturally falls into this
general form, where $\F(\partial_{x}, u)=-(uu_x+u_{xx}+\gamma
u_{xxxx})$, which does not depend explicitly on $x$. The
damped-forced NLS equation (\ref{e:DDNLS}) falls into this general
form as well when it is rewritten in terms of the real and imaginary
parts of the complex function $\psi$ (which make up the real vector
variable $\u$).

Since Eq. (\ref{e:ut1}) admits a time-periodic solution $\u(\x, t)$
with an unknown temporal period $T$, i.e., $\u(\x, t+T)=\u(\x, t)$,
it proves convenient to introduce a time scaling
\[ \label{e:tau}
\tau=\omega t, \quad \omega\equiv 2\pi/T.
\]
Under this scaling, Eq. (\ref{e:ut1}) becomes
\[ \label{e:utau1}
\omega \u_\tau=\F(\x, \partial_{\x}, \u),
\]
where $\u(\x, \tau)$ is $2\pi$-periodic in $\tau$, i.e.,
\[  \label{e:utau1period}
\u(\x, \tau+2\pi)=\u(\x, \tau).
\]
Thus the computational domain for $\u(\x, \tau)$ can be set
explicitly as $0\le \tau \le 2\pi$ and $\x\in \Omega$, where
$\Omega$ is the $\x$-domain of the solution $\u(\x, t)$. In the
scaled equation (\ref{e:utau1}), the frequency $\omega$ is the new
unknown parameter.

To solve Eq. (\ref{e:utau1}) with the temporal periodicity condition
(\ref{e:utau1period}) and unknown frequency $\omega$, our idea is to
first express this unknown frequency $\omega$ in terms of the
periodic solution $\u(\x, \tau)$ through a Rayleigh-like quotient.
That is, we take the inner product of Eq. (\ref{e:utau1}) with
function $\u_\tau$, and then obtain $\omega$ as
\[  \label{e:omega}
\omega=\frac{\langle \u_\tau, \F \rangle} {\langle \u_\tau, \u_\tau \rangle}.
\]
Here the inner product is the standard one in the real-valued vector
functional space
\[  \label{e:innprod}
\langle \textbf{f}(\x, \tau), \textbf{g}(\x, \tau) \rangle =\int_0^{2\pi}\int_\Omega  \textbf{f}^{\mbox{\scriptsize T}} \textbf{g} \; \rd \x \, \rd\tau,
\]
where the superscript `\mbox{\scriptsize T}' represents transpose of
a vector. In this article, we call the Rayleigh-like quotient
(\ref{e:omega}) as a quasi-Rayleigh quotient. Inserting this
quasi-Rayleigh quotient (\ref{e:omega}) into (\ref{e:utau1}), we
then get the equation
\[ \label{e:L0u}
\L_0(\u)\equiv  \frac{\langle \u_\tau, \F \rangle} {\langle \u_\tau, \u_\tau
\rangle} \u_\tau - \F=0.
\]
In this equation, the unknown frequency $\omega$ is gone, thus the
equation is for the unknown function $\u(\x, \tau)$ only. The price
to pay for this benefit is that this equation now becomes an
integro-differential equation instead of a differential equation.
But this is a price worth paying for.

We solve the integro-differential equation (\ref{e:L0u}) by the
Newton-conjugate-gradient (Newton-CG) method \cite{Yang2010}. In
this method, conjugate gradient (CG) iterations are used to solve
the linear Newton-correction equation. Suppose $\u_n(\x, \tau)$ is
the $n$-th approximation to the exact solution, then the Newton
iteration for the next approximation is
\[
\u_{n+1}=\u_n + \Delta \u_n,
\]
where the linear Newton-correction equation for $\Delta \u_n$ is
\[  \label{e:Newtoncorrection}
\L_{1n} \Delta \u_n = -\L_0(\u_n).
\]
Here $\L_1$ is the linearization operator of function $\L_0(\u)$,
i.e.,
\[
\L_0(\u+\tilde{\u})=\L_0(\u)+\L_1\tilde{\u}+O(\tilde{\u}^2), \quad \tilde{\u} \ll 1,
\]
and $\L_{1n}$ is $\L_1$ evaluated at $\u=\u_n$. This linearization
operator $\L_1$ is the counterpart of the Jacobian in systems of
nonlinear ordinary differential equations.

Now we derive the analytical expression for $\L_1$. Suppose the
linearization operator for the function $\F(\x, \partial_{\x}, \u)$
is $\F_1$, i.e.,
\[ \label{e:Flinearize}
\F(\x, \partial_{\x}, \, \u+\tilde{\u})=\F(\x, \partial_{\x}, \u)+\F_1\tilde{\u}+O(\tilde{\u}^2), \quad \tilde{\u} \ll 1.
\]
When a dissipative wave system (\ref{e:ut1}) is given, the function
$\F(\x, \partial_{\x}, \u)$ is known, thus its linearization
operator $\F_1$ can be analytically obtained (this calculation for
the Kuramoto-Sivashinsky equation (\ref{e:KS}) will be demonstrated
in section \ref{e:secnum}). Using the linearization
(\ref{e:Flinearize}) for $\F$, the linearization for $\omega(\u)$ in
Eq. (\ref{e:omega}) is
\begin{equation*}
\omega(\u+\tilde{\u}) = \frac{\langle (\u+\tilde{\u})_\tau, \, \F(\x, \partial_{\x}, \, \u+\tilde{\u}) \rangle} {\langle (\u+\tilde{\u})_\tau, (\u+\tilde{\u})_\tau \rangle}
=\frac{\langle (\u+\tilde{\u})_\tau, \, \F(\x, \partial_{\x}, \u)+\F_1\tilde{\u} \rangle} {\langle (\u+\tilde{\u})_\tau, (\u+\tilde{\u})_\tau \rangle}+O(\tilde{\u}^2).
\end{equation*}
Utilizing Eq. (\ref{e:utau1}), we get
\[
\omega(\u+\tilde{\u})=\frac{\langle (\u+\tilde{\u})_\tau, \, \omega(\u) \u_\tau+\F_1\tilde{\u} \rangle} {\langle (\u+\tilde{\u})_\tau, (\u+\tilde{\u})_\tau \rangle}+O(\tilde{\u}^2)
=\frac{\langle (\u+\tilde{\u})_\tau, \, \omega(\u) (\u+\tilde{\u})_\tau-[\omega(\u)\partial_\tau-\F_1]\tilde{\u} \rangle} {\langle (\u+\tilde{\u})_\tau, (\u+\tilde{\u})_\tau \rangle}+O(\tilde{\u}^2),
\nonumber
\]
thus
\[
\omega(\u+\tilde{\u})=\omega(\u)-\frac{\langle \u_\tau, \, [\omega(\u)\partial_\tau-\F_1]\tilde{\u} \rangle} {\langle \u_\tau, \u_\tau \rangle}+O(\tilde{\u}^2).
\]
Using this $\omega(\u)$ linearization as well as the $\F(\x,
\partial_{\x}, \u)$ linearization (\ref{e:Flinearize}), the
linearization operator $\L_1$ for $\L_0(\u)$ can then be found as
\[  \label{e:L1}
\L_1\Psi=\P\Psi-\frac{\langle \u_\tau, \, \P\Psi \rangle} {\langle \u_\tau, \u_\tau \rangle}\u_\tau,
\]
where
\[
\P \equiv \omega \partial_\tau-\F_1,
\]
and $\omega(\u)$ is given through $\u$ by the quasi-Rayleigh
quotient (\ref{e:omega}).

It is now time to discuss how to solve the linear Newton-correction
equation (\ref{e:Newtoncorrection}). In the Newton-CG method, this
equation will be solved by conjugate-gradient iterations, which is
widely recognized as probably the fastest way to solve large systems
of linear inhomogeneous equations \cite{Golub}. Since the
homogeneous operator $\L_1$ in (\ref{e:L1}) is apparently
non-self-adjoint, it is necessary to turn equation
(\ref{e:Newtoncorrection}) into a sort of normal equation so that
its homogeneous operator becomes self-adjoint. The usual way to turn
(\ref{e:Newtoncorrection}) into a normal equation is to multiply it
by the adjoint operator of $\L_{1}$. But due to the special
structure of $\L_{1}$ in (\ref{e:L1}), we can ``cut corners" and
just multiply (\ref{e:Newtoncorrection}) by the adjoint operator of
$\P$, which is
\[  \label{e:PA}
\P^A=-\omega \partial_\tau-\F_1^A,
\]
where $\F_1^A$ is the adjoint operator of $\F_1$. Here the
superscript `{\small $A$}' represents the adjoint. With this
multiplication, the Newton-correction equation
(\ref{e:Newtoncorrection}) becomes
\[  \label{e:normal}
\P_n^A \L_{1n} \Delta \u_n = -\P_n^A \L_0(\u_n),
\]
where $\P_n^A$ is $\P^A$ evaluated at $\u=\u_n$. For convenience, we
call this equation a quasi-normal equation. It is easy to check that
$\P^A\L_1$ is self-adjoint. In addition, using the Cauchy-Schwarz
inequality, we can show that $\P^A\L_1$ is also semi-positive
definite. Thus the quasi-normal equation (\ref{e:normal}) can be
solved by preconditioned conjugate gradient iterations. The
numerical algorithm for preconditioned conjugate gradient iterations
is well known \cite{Golub} and will not be repeated here (the reader
can refer to the sample MATLAB codes in the appendices for numerical
executions of these iterations). We do want to mention that, in
order to avoid over-solving, CG iterations for the quasi-normal
equation (\ref{e:normal}) will be stopped when the error of the
Newton-correction solution $\Delta \u_n$ drops below a certain
fraction of the error of the solution $\u_n$ itself \cite{YangCG}
(in our coding, this fraction is set as \verb|errorCG|$=10^{-4}$,
see appendices). This strategy reduces the number of CG iterations
for solving each Newton-correction equation at the expense of losing
the quadratic convergence of Newton iterations, but its benefit
outweighs its price \cite{YangCG}.

To summarize, our numerical algorithm for computing time-periodic
solutions $\u(\x, t)$ with unknown temporal periods in Eq.
(\ref{e:ut1}) is:
\begin{enumerate}
\item turn (\ref{e:ut1}) into an integro-differential equation (\ref{e:L0u}) for the function $\u(\x, \tau)$
only, under the time-periodic boundary condition
(\ref{e:utau1period});
\item solve (\ref{e:L0u}) by Newton iterations
\[
\u_{n+1}=\u_n + \Delta \u_n,   \nonumber
\]
where Newton corrections $\Delta\u_n$ are computed from the
quasi-normal equation
\[
\P_n^A \L_{1n} \Delta \u_n = -\P_n^A \L_0(\u_n)  \nonumber
\]
by preconditioned conjugate gradient iterations. Here, linear
operators $\L_1$ and $\P^A$ are given analytically by equations
(\ref{e:L1})-(\ref{e:PA});
\item after $\u(\x, \tau)$ is obtained, the temporal period $T (=2\pi/\omega)$ is
then derived from the quasi-Rayleigh quotient (\ref{e:omega}).
\end{enumerate}

The above numerical algorithm is attractive for a number of reasons.
First the entire computation is performed in the PDE framework (no
truncation to ODEs or algebraic equations is necessary). Second, it
is applicable to both stable and unstable time-periodic solutions.
This contrasts the time-evolution method which can only converge to
stable solutions. Thirdly, its numerical accuracy can be very high.
Indeed, if we use the discrete Fourier transform or Chebyshev
differentiation to compute all spatial and temporal derivatives,
then its numerical accuracy would be spectral
\cite{Trefethen2000,Boyd2001}. Fourthly, this method is
fast-converging and very efficient. This efficiency will be
illustrated on several numerical examples in section \ref{e:secnum},
where we will see that this method only takes from a fraction of a
second to a couple of minutes (on a personal computer) to find
solutions of varying complexities to the accuracy of $10^{-10}$.
Fifthly, the coding of this method is very short and compact, as is
evidenced in the sample MATLAB codes to be presented in the
appendices.

In the implementation of the above Newton-CG method, there are two
practical issues. One is the choice of the preconditioning operator
for solving the quasi-normal equation by preconditioned conjugate
gradient iterations. This preconditioner, say $\M$, must be
self-adjoint and positive definite. In addition, it should make the
condition number of $\M^{-1}\P^A\L_1$ as small as possible (i.e., to
make $\M^{-1}\P^A\L_1$ as close to the identity operator as
possible) in order to get faster convergence. Furthermore it should
be easy to invert, since this inversion is needed during iterations.
Since the large condition number of $\P^A\L_1$ in the quasi-normal
equation, which slows down CG iterations, is generally caused by
higher space and time derivatives in $\P^A\L_1$, then a general
guideline for the choice of the preconditioner is to retain only the
higher-derivative terms in $\P^A\L_1$ and use the resulting operator
as $\M$ (added by a positive constant to make $\M$
positive-definite). Implementation of this guideline on several
numerical examples will be illustrated in section \ref{e:secnum}.

The other practical issue in the Newton-CG method is the choice of
the initial condition. It is well known that if the initial
condition is not properly chosen, Newton iterations may not
converge. There are various strategies for choosing the initial
condition. The first strategy is to just choose the initial
condition randomly. This strategy may work, especially if the
solution has a simple structure, but one often needs to try many
initial conditions in order to hit upon one that works. A second
strategy is to simulate the time evolution of the original wave
equation and inspect the solution to see if any time-segment of this
solution is close to time-periodic or not. If so, then that
time-segment of the solution will be used as our initial condition
for Newton-CG iterations. Note that this second strategy is
applicable to both stable and unstable time-periodic solutions,
since even if the solution is unstable, time evolution of the wave
equation may still get close to this solution and wander around it
for a little while (before veering off), and that approximate
time-periodic segment is often sufficient as our initial condition
for Newton-CG iterations. A third strategy is by continuation. If we
have obtained a time-periodic solution at one parameter value, then
by continuously changing this parameter and using the previous
solution as the initial condition, we can trace a whole family of
time-periodic solutions for a continuous range of this parameter.
This continuation strategy is often very useful, especially for
studying bifurcations of solutions as parameters vary. In our
numerical examples of section \ref{e:secnum}, we will apply all
these strategies to select initial conditions of Newton-CG
iterations for both stable and unstable time-periodic solutions.

\section{A numerical method for time-periodic solutions with multiple unknown parameters}

In some dissipative wave systems, time-periodic solutions have more
unknown parameters than just the temporal period. One example is the
Ginzburg-Landau-type equations such as
\[  \label{e:CQGL}
A_t=\chi A+\gamma A_{xx}-\beta |A|^2A-\delta |A|^4A,
\]
where $A$ is a complex variable, and $\chi, \gamma, \beta, \delta$
are complex coefficients. This equation does not admit truly
time-periodic solutions, but it admits the so-called relative
time-periodic solutions of the form
\[ \label{e:A}
A(x, t)=e^{i\mu t}U(x, t),
\]
where $U(x, t)$ is a time-periodic complex function, and $\mu$ is a
real-valued propagation constant
\cite{Deissler1994,Akhmediev2000,Lopez2005}. In this solution, both
the temporal period $T$ of $U(x, t)$ and the propagation constant
$\mu$ are unknown in addition to the unknown function $U(x, t)$. In
order to compute these relative time-periodic solutions, the
numerical algorithm of the previous section needs to be modified and
generalized.

In this section, we develop a numerical scheme to compute
time-periodic solutions with multiple unknown parameters (here
`time-periodic solutions' is interpreted in the broader sense,
including relative time-periodic solutions). The basic idea is
similar to that of the previous section. That is, we first express
these unknown parameters in terms of the time-periodic function
through quasi-Rayleigh quotients so that the original wave equation
becomes an integro-differential equation for the unknown
time-periodic function only. Then we use Newton-CG iterations to
solve this integro-differential equation. But since the current
problem involves multiple unknown parameters, the linearization
operator of the integro-differential equation will have a different
structure than Eq. (\ref{e:L1}) of the previous section. Because of
that, we will have to solve the linear Newton-correction equation
(through CG iterations) by turning it into a true normal equation
instead of a quasi-normal equation. That is, we will need to
multiply the Newton-correction equation by the adjoint of the whole
Newton-linearization operator rather than a partial one.

Even though our basic idea for computing time-periodic solutions
with multiple unknown parameters is easy to state, formulation of
this idea for general dissipative systems can be cumbersome. Thus in
the following, we only formulate this idea for a special (but
important) class of equations, namely the Ginzburg-Landau-type
equations. Extension of this formulation to other types of equations
is straightforward.

The class of Ginzburg-Landau-type equations that we consider can be
written in the following general form,
\[ \label{e:AGL}
A_t=f(|A|, \x, \partial_{\x})A,
\]
where $\x=(x_1, x_2, \cdots, x_N)$ is the $N$-dimensional spatial
coordinate, $A(\x, t)$ is a complex-valued scalar variable of $\x$
and time $t$, and $f$ is a complex-valued function of $|A|$, $\x$
and the spatial derivatives. As before, we allow $f$ to contain
explicit dependence on $\x$ (to model spatial inhomogeneities), but
not on time $t$. The cubic-quintic Ginzburg-Landau equation
(\ref{e:CQGL}) is an example of this general form, with $f(|A|,
\partial_{x})=\chi +\gamma
\partial_{xx}-\beta |A|^2-\delta |A|^4$, which contains no explicit
$x$-dependence.

This class of Ginzburg-Landau-type equations admit relative
time-periodic solutions of the form
\[ \label{e:A2}
A(\x, t)=e^{i\mu t}U(\x, t),
\]
where $U(\x, t)$ is a time-periodic complex function, and $\mu$ is a
real-valued propagation constant. These solutions can be spatially
localized or periodic \cite{Deissler1994,Akhmediev2000,Lopez2005}.
Both the temporal period $T$ and the propagation constant $\mu$ are
not known priori and must be determined along with the time-periodic
function $U(\x, t)$.

Substituting (\ref{e:A2}) into Eq. (\ref{e:AGL}), we get the
equation for the time-periodic function $U(\x, t)$ as
\[ \label{e:U}
U_t+i\mu U=G,
\]
where
\[ \label{def:G}
G\equiv f(|U|, \x, \partial_{\x})U.
\]
As before, we employ a time scaling
\[ \label{e:tau2}
\tau=\omega t, \quad \omega\equiv 2\pi/T,
\]
where $T$ is the temporal period of the function $U(\x, t)$. Under
this scaling, Eq. (\ref{e:U}) becomes
\[ \label{e:Utau}
\omega U_\tau+i\mu U= G,
\]
where $U(\x, \tau)$ is $2\pi$-periodic in $\tau$, i.e.,
\[  \label{e:Utau1period}
U(\x, \tau+2\pi)=U(\x, \tau).
\]
Thus the computational domain for $U(\x, \tau)$ will be set
explicitly as $0\le \tau \le 2\pi$ and $\x\in \Omega$, where
$\Omega$ is the $\x$-domain of the solution $U(\x, t)$.

To solve Eq. (\ref{e:Utau}), we first express the unknown real
parameters $\omega$ and $\mu$ in terms of $U(\x, \tau)$ through
quasi-Rayleigh quotients. For this purpose, it is convenient to
split the complex functions $U$ and $G$ into real and imaginary
parts as
\[
U=u+iv, \quad G=g+ih.
\]
Inserting this split into Eq. (\ref{e:Utau}), equations for the real
and imaginary parts $u, v$ of the solution can be readily obtained
as
\begin{eqnarray}
\omega u_{\tau}-\mu v -g=0,  \label{e:u1tau}\\
\omega v_{\tau}+\mu u -h=0.  \label{e:u2tau}
\end{eqnarray}
Taking inner products of these equations with $u, v, u_{\tau},
v_{\tau}$, adding or subtracting the resulting equations, and
utilizing the $\tau$-periodicity of $(u, v)$, parameters $\mu$ and
$\omega$ can be expressed through the following quasi-Rayleigh
quotients,
\[  \label{e:muomega}
\mu=\frac{\langle v, h\rangle-\langle u, g\rangle}{2\langle u, v\rangle},  \qquad
\omega=\frac{\langle u_{\tau}, h\rangle + \langle v_{\tau}, g\rangle }{2\langle u_{\tau}, v_{\tau}\rangle}.
\]
Here the inner product is the same as that defined in Eq.
(\ref{e:innprod}). Inserting these quasi-Rayleigh quotients into
(\ref{e:u1tau})-(\ref{e:u2tau}), these equations then become the
following integro-differential equations for the unknown functions
$\u \equiv [u, v]^{\mbox{\scriptsize T}}$ only,
\[ \label{e:L0u1u2}
\L_0(\u)\equiv \left[ \begin{array}{c} \omega u_{\tau}-\mu v -g \\
\omega v_{\tau}+\mu u -h \end{array}\right]=0,
\]
where $\mu(\u)$ and $\omega(\u)$ are given in equation
(\ref{e:muomega}).

We use Newton-CG methods to solve the integro-differential equations
(\ref{e:L0u1u2}). As before, the Newton iterations are
\[
\u_{n+1}=\u_n + \Delta \u_n,
\]
where the linear Newton-correction equation for $\Delta \u_n$ is
\[  \label{e:Newtoncorrection2}
\L_{1n} \Delta \u_n = -\L_0(\u_n),
\]
and $\L_1$ is the linearization operator of function $\L_0(\u)$. The
key question now is the analytical expression for this linearization
operator, which is certainly different from (\ref{e:L1}) of the
previous section. This expression of $\L_1$ is given in the
following lemma.

\vspace{0.1cm} \textbf{Lemma~1} \hspace{0.1cm} The linearization
operator $\L_1$ of $\L_0(\u)$ in Eq. (\ref{e:L0u1u2}) is
\[  \label{e:L1formula}
\L_1\Psi=\P\Psi - \frac{\left\langle \left[\begin{array}{c} v_\tau \\ u_\tau \end{array}\right], \P\Psi\right\rangle}{2\langle u_\tau, v_\tau\rangle}
\left[\begin{array}{c} u_\tau \\ v_\tau \end{array}\right]+
\frac{\left\langle \left[\begin{array}{c} u \\ -v \end{array}\right], \P\Psi\right\rangle}{2\langle u, v\rangle}
\left[\begin{array}{c} -v \\ u \end{array}\right],
\]
where
\[  \label{e:defP}
\P=\left[\begin{array}{cc} \omega\partial_\tau & -\mu \\  \mu & \omega\partial_\tau\end{array}\right]-\G_1,
\]
and $\G_1$ is the linearization operator of the vector function $[g,
h]^{\mbox{\scriptsize T}}$, i.e.,
\[  \label{e:ghlinearize}
\left[\begin{array}{c} g(u+\tilde u, v+\tilde v) \\ h(u+\tilde u, v+\tilde v) \end{array}\right]=\left[\begin{array}{c} g(u,v) \\ h(u,v)
\end{array}\right]+\G_1 \left[\begin{array}{c} \tilde u \\
\tilde v \end{array}\right] +O(\tilde{u}^2, \tilde{u}\tilde{v}, \tilde{v}^2).
\]

\vspace{0.1cm} The proof of this lemma will be provided later in
this section. For the example of the cubic-quintic Ginzburg-Landau
equation (\ref{e:CQGL}), calculation of the linear operator $\G_1$
will be illustrated in section \ref{e:secnum}.

We can notice that the linearization operator $\L_1$ in this lemma
has a more complex structure than that in (\ref{e:L1}) of the
previous section. Because of that, in order to turn $\L_1$ into a
self-adjoint operator, we have to multiply it by its full adjoint
$\L_1^A$. In other words, in order to solve the linear
Newton-correction equation (\ref{e:Newtoncorrection2}) by conjugate
gradient iterations, we need to turn it into the usual normal
equation
\[  \label{e:normal2}
\L_{1n}^A \L_{1n} \Delta \u_n = -\L_{1n}^A \L_0(\u_n).
\]
Compared with the previous quasi-normal equation (\ref{e:normal}),
we have no ``corners to cut" here. Obviously the linear operator
$\L_{1n}^A \L_{1n}$ in the above normal equation is self-adjoint and
semi-positive definite, thus this equation can be solved effectively
by preconditioned conjugate gradient iterations.

The normal equation (\ref{e:normal2}) involves the adjoint operator
$\L_{1}^A$. This adjoint operator can be derived from $\L_1$ in
Lemma 1, and its analytical expression is provided by the following
lemma.

\vspace{0.1cm} \textbf{Lemma~2} \hspace{0.1cm} The adjoint operator
of $\L_1$ in Lemma 1 is
\[ \label{e:L1Adef}
\L_1^A\Psi=\P^A\Psi - \frac{\left\langle \Psi, \left[\begin{array}{c} u_\tau \\ v_\tau \end{array}\right] \right\rangle} {2\langle u_\tau, v_\tau\rangle}
\P^A\left[\begin{array}{c} v_\tau \\ u_\tau \end{array}\right]+
\frac{\left\langle \Psi, \left[\begin{array}{c} -v \\ u \end{array}\right] \right\rangle} {2\langle u, v\rangle} \P^A \left[\begin{array}{c} u
\\ -v \end{array}\right],
\]
where
\[  \label{e:PA2}
\P^A=\left[\begin{array}{cc} -\omega\partial_\tau & \mu \\  -\mu & -\omega\partial_\tau\end{array}\right]-\G_1^A
\]
is the adjoint operator of $\P$, and $\G_1^A$ is the adjoint
operator of $\G_1$.

\vspace{0.1cm} The proof for this lemma will follow shortly.

To summarize, our numerical algorithm for computing relative
time-periodic solutions (\ref{e:A2}), with unknown temporal period
$T$ and propagation constant $\mu$, in the class of
Ginzburg-Landau-type equations (\ref{e:AGL}) is
\begin{enumerate}
\item turn (\ref{e:AGL}) into an integro-differential equation (\ref{e:L0u1u2}) for the real and imaginary parts $\u \equiv
[u, v]^{\mbox{\scriptsize T}}$ of the solution $U(\x, \tau)$
only, under the time-periodic boundary condition
(\ref{e:Utau1period});
\item solve (\ref{e:L0u1u2}) by Newton iterations
\[
\u_{n+1}=\u_n + \Delta \u_n,   \nonumber
\]
where Newton corrections $\Delta\u_n$ are computed from the
normal equation
\[
\L_{1n}^A \L_{1n} \Delta \u_n = -\L_{1n}^A \L_0(\u_n)   \nonumber
\]
by preconditioned conjugate gradient iterations. Here, linear
operators $\L_1$ and $\L_1^A$ are given analytically in Lemmas~1
and 2.
\item after $\u(\x, \tau)$ is obtained, the temporal period $T (=2\pi/\omega)$ and the propagation constant $\mu$
are then calculated from the quasi-Rayleigh quotients
(\ref{e:muomega}).

\end{enumerate}

This numerical method shares the same attractive features as that
described in the previous section (such as high accuracy,
efficiency, short coding, and so on). In the implementation of this
method, we also face the two practical issues discussed in the end
of section 2, which are choices of the preconditioner and the
initial condition. Our guidelines for these choices are the same as
those spelled out there.

Now we prove Lammas 1 and 2.

\vspace{0.1cm} \textbf{Proof of Lemma 1.}  \hspace{0.1cm} We first
derive linearizations for the quasi-Rayleigh quotients of $\mu$ and
$\omega$. For this purpose, we rewrite the $\mu$ formula as
\[
\mu(u, v)=\frac{\left\langle \left[\begin{array}{c} -u \\ v \end{array}\right], \,  \left[\begin{array}{c} g(u, v) \\ h(u, v) \end{array}\right]\right\rangle}
{2\langle u, v \rangle}.   \nonumber
\]
Utilizing the linearization (\ref{e:ghlinearize}) for $[g,
h]^{\mbox{\scriptsize T}}$, we get
\[
\mu(u+\tilde{u}, v+\tilde v)=\frac{\left\langle \left[\begin{array}{c} -(u+\tilde u) \\ v+\tilde v \end{array}\right], \,
\left[\begin{array}{c} g(u, v) \\ h(u, v) \end{array}\right] + \G_1 \left[\begin{array}{c} \tilde u \\ \tilde v \end{array}\right] \right\rangle}
{2\langle u+\tilde u, v+\tilde v \rangle}+O(\tilde{u}^2, \tilde{u}\tilde{v}, \tilde{v}^2).  \nonumber
\]
Then using equations (\ref{e:u1tau})-(\ref{e:u2tau}), we can
calculate $\mu(u+\tilde{u}, v+\tilde v)$ as
\begin{eqnarray*}
\mu(u+\tilde{u}, v+\tilde v)& = & \frac{\left\langle \left[\begin{array}{c} -(u+\tilde u) \\ v+\tilde v \end{array}\right], \,\,
\left[\begin{array}{c} \omega u_\tau-\mu v \\ \omega v_\tau+\mu u \end{array}\right] + \G_1 \left[\begin{array}{c} \tilde u \\ \tilde v \end{array}\right] \right\rangle}
{2\langle u+\tilde u, v+\tilde v \rangle}+O(\tilde{u}^2, \tilde{u}\tilde{v}, \tilde{v}^2) \\
&=& \frac{\left\langle \left[\begin{array}{c} -(u+\tilde u) \\ v+\tilde v \end{array}\right], \,\,
\omega \left[\begin{array}{c} u+\tilde u \\ v+\tilde v\end{array}\right]_\tau +
\mu \left[\begin{array}{c} -(v+\tilde v) \\ u+\tilde u \end{array}\right]
- \P \left[\begin{array}{c} \tilde u \\ \tilde v \end{array}\right] \right\rangle}
{2\langle u+\tilde u, v+\tilde v \rangle}+O(\tilde{u}^2, \tilde{u}\tilde{v}, \tilde{v}^2),
\end{eqnarray*}
where $\mu, \omega$ on the right sides of these equations are
abbreviations for $\mu(u, v), \omega(u, v)$, and operator $\P$ is as
defined in (\ref{e:defP}). Recalling the $\tau$-periodicity of
functions $u+\tilde u$ and $v+\tilde v$, the above expression for
$\mu(u+\tilde{u}, v+\tilde v)$ can be simplified as
\[
\mu(u+\tilde{u}, v+\tilde v)=\mu(u, v)-
\frac{\left\langle \left[\begin{array}{c} -u \\ v \end{array}\right], \,
\P \left[\begin{array}{c} \tilde u \\ \tilde v \end{array}\right] \right\rangle}
{2\langle u, v\rangle}+O(\tilde{u}^2, \tilde{u}\tilde{v}, \tilde{v}^2),  \nonumber
\]
which is the linearization for the quasi-Rayleigh quotient of $\mu$.
Performing similar calculations, the linearization for the
quasi-Rayleigh quotient of $\omega$ is found as
\[
\omega(u+\tilde{u}, v+\tilde v)=\omega(u, v)-
\frac{\left\langle \left[\begin{array}{c} v_\tau \\ u_\tau \end{array}\right], \,
\P \left[\begin{array}{c} \tilde u \\ \tilde v \end{array}\right] \right\rangle}
{2\langle u_\tau, v_\tau\rangle}+O(\tilde{u}^2, \tilde{u}\tilde{v}, \tilde{v}^2).  \nonumber
\]
Using these linearizations of $\mu$ and $\omega$ as well as the
linearization (\ref{e:ghlinearize}) of $[g, h]^{\mbox{\scriptsize
T}}$, the linearization operator $\L_1$ for equation
(\ref{e:L0u1u2}) is then found to be (\ref{e:L1formula}) in Lemma 1.

\vspace{0.15cm} \textbf{Proof of Lemma 2.}  \hspace{0.1cm}  The
definition for adjoint operators is that
\[
\langle \Phi, \L_1\Psi\rangle = \langle \L_1^A\Phi, \Psi\rangle.   \nonumber
\]
Using the expression of $\L_1$ in Lemma 1 as well as the basic
relation of $\langle \Phi, \P \Psi\rangle = \langle \P^A\Phi,
\Psi\rangle$, we find that
\begin{eqnarray*}
\langle \Phi, \L_1\Psi\rangle & = & \langle \Phi, \P\Psi\rangle -
\frac{\left\langle \left[\begin{array}{c} v_\tau \\ u_\tau \end{array}\right], \P\Psi\right\rangle
\left\langle \Phi, \left[\begin{array}{c} u_\tau \\ v_\tau \end{array}\right]\right\rangle}
{2\langle u_\tau, v_\tau\rangle}+
\frac{\left\langle \left[\begin{array}{c} u \\ -v \end{array}\right], \P\Psi\right\rangle
\left\langle \Phi,  \left[\begin{array}{c} -v \\ u
\end{array}\right]\right\rangle}{2\langle u, v\rangle} \\
& = &
\langle \P^A \Phi, \Psi\rangle -
\frac{\left\langle \Phi, \left[\begin{array}{c} u_\tau \\ v_\tau \end{array}\right]\right\rangle
\left\langle \P^A \left[\begin{array}{c} v_\tau \\ u_\tau \end{array}\right], \Psi\right\rangle}
{2\langle u_\tau, v_\tau\rangle}+
\frac{\left\langle \Phi,  \left[\begin{array}{c} -v \\ u
\end{array}\right]\right\rangle \left\langle \P^A\left[\begin{array}{c} u \\ -v \end{array}\right], \Psi\right\rangle
}{2\langle u, v\rangle},
\end{eqnarray*}
which is the same as $\langle \L_1^A\Phi, \Psi\rangle$ with $\L_1^A$
given in Lemma 2.

To prove $\P^A$ in equation (\ref{e:PA2}) is the adjoint operator of
$\P$, we only need to use the definition of adjoint operators,
together with integration by parts and the fact that all admissible
functions are $\tau$-periodic.

Before concluding this section, we would like to make a remark. As
the reader may notice, the expressions of $\mu$ and $\omega$ through
quasi-Rayleigh quotients are not unique. Indeed, from equations
(\ref{e:u1tau})-(\ref{e:u2tau}) we can also derive other
quasi-Rayleigh quotients of $\mu$ and $\omega$ different from
(\ref{e:muomega}). For instance, by taking the inner products of
(\ref{e:u1tau}) with $u$ and (\ref{e:u2tau}) with $u_\tau$, and
utilizing the $\tau$-periodicity of the involved functions, we can
obtain the following alternative expressions
\[
\mu=-\frac{\langle u, g\rangle}{\langle u, v\rangle}, \quad
\omega=\frac{\langle u_\tau, h\rangle}{\langle u_\tau, v_\tau\rangle}.  \nonumber
\]
Substituting these alternative quasi-Rayleigh quotients into
equations (\ref{e:u1tau})-(\ref{e:u2tau}), we can still use
Newton-CG iterations to solve them, except that the linearization
operator $\L_1$ and its adjoint $\L_1^A$ will be different from
those in Lemmas~1 and 2. We have implemented this and several other
versions of quasi-Rayleigh quotients on the cubic-quintic
Ginzburg-Landau equation (\ref{e:CQGL}), and found that their
performances are slightly inferior to the quasi-Rayleigh quotients
in equation (\ref{e:muomega}). The reason is probably that $\mu$ and
$\omega$ in formulae (\ref{e:muomega}) are derived by taking the
average of inner products from equations (\ref{e:u1tau}) and
(\ref{e:u2tau}). This averaging may give more accurate
approximations for $\mu$ and $\omega$ from an approximate solution
$(u_n, v_n)$, thus rendering the numerical scheme superior to some
other alternatives.

\section{Numerical examples}  \label{e:secnum}

In this section, we apply the proposed numerical methods of previous
sections to three well-known dissipative systems, the
Kuramoto-Sivashinsky equation, the cubic-quintic Ginzburg-Landau
equation, and the Lorenz equations. Both stable and unstable
time-periodic solutions in these equations will be computed. All our
computations are performed in MATLAB on a Desktop PC (Dell Optiplex
990 with CPU speed 3.3GHz). MATLAB codes for these computations can
be found in the appendices.

\vspace{0.1cm} \textbf{Example 1} \; Our first example is the
Kuramoto-Sivashinsky (KS) equation (\ref{e:KS}), i.e.,
\[  \label{e:KS2}
u_t+uu_x+u_{xx}+\gamma u_{xxxx}=0,
\]
where $u$ is a scalar real variable, and $\gamma$ is a real
``superviscosity" coefficient. This equation was derived in various
physical contexts as a model for wave dynamics near long-wave-length
instabilities in the presence of certain symmetries
\cite{LaQuey1975,Kuramoto1976,Sivashinsky1977}. But it is also used
to study spatiotemporal complexity
\cite{Hyman1986,Kevrekidis1990,Papageorgiou1991,Zoldi1998,Cvitanovic2004}.
In these studies, it is customary to impose the periodic boundary
condition
\[
u(x+2\pi, t)=u(x, t).
\]
Under this boundary condition, we seek time-periodic solutions in
this equation. Thus the numerical domain of our algorithm will be
set as $0\le x, \tau  \le 2\pi$. Since the only unknown parameter in
these solutions is the temporal period, the algorithm in section 2
will be suitable. This algorithm is capable of obtaining both stable
and unstable time-periodic solutions. Below we will apply it to
determine unstable solutions, since such solutions cannot be
obtained by the time-evolution method and are thus more challenging
to find.

For the KS equation (\ref{e:KS2}), the function $\F$ in the
algorithm of section 2 is
\[
\F(\partial_{x}, u)=-(uu_x+u_{xx}+\gamma
u_{xxxx}).    \nonumber
\]
Linearization of this function is
\begin{eqnarray*}
\F(\partial_x, u+\tilde{u}) & = &-\left[(u+\tilde u) (u+\tilde u)_x+(u+\tilde u)_{xx}+\gamma (u+\tilde u)_{xxxx}\right] \\
&=&  \F(\partial_x, u) - \left[ u\partial_x +u_x+\partial_{xx}+\gamma\partial_{xxxx}\right] \tilde u + O(\tilde{u}^2),
\end{eqnarray*}
thus the linearization operator of $\F$ is
\[
\F_1=- \left[ u\partial_x +u_x+\partial_{xx}+\gamma\partial_{xxxx}\right].   \nonumber
\]
Its adjoint operator $\F_1^A$ can be easily derived from the basic
condition $\langle \phi, \F_1\psi\rangle = \langle \F_1^A \phi,
\psi\rangle$ as
\[
\F_1^A = -\left[ -u\partial_x + \partial_{xx}+\gamma\partial_{xxxx}\right].   \nonumber
\]
Using these formulae, the quasi-normal Newton-correction equation
(\ref{e:normal}) for the KS equation is
\[  \label{e:normalKS}
\P_n^A \L_{1n} \Delta u_n = -\P_n^A \L_0(u_n),
\]
where
\[
\L_0(u)=\omega u_\tau-\F, \quad \L_1\psi\equiv
\P\psi-\frac{\langle u_\tau, \, \P\psi \rangle} {\langle u_\tau, u_\tau \rangle}u_\tau, \quad \P=\omega \partial_\tau-\F_1, \quad
\P^A=-\omega \partial_\tau-\F_1^A, \quad \omega=\frac{\langle u_\tau, \F \rangle} {\langle u_\tau, u_\tau \rangle},  \nonumber
\]
and the quantities with subscript `$n$' in (\ref{e:normalKS}) are
the corresponding quantities evaluated at the $n$-th approximate
solution $u_n$.

Regarding the preconditioner $\M$ in preconditioned conjugate
gradient iterations on the quasi-normal equation (\ref{e:normalKS}),
we take
\[
\M=c-\omega^2\partial_{\tau\tau}+(\partial_{xx}+\gamma\partial_{xxxx})^2,   \nonumber
\]
where $c$ is a positive constant (which we choose as $c=30$; other
$c$ values deliver comparable performances). Our choice of this
preconditioner follows the guidelines at the end of section 2.
Specifically, neglecting lower-derivative terms in $\P^A\L_1$, we
get
\[
\P^A\L_1 \approx -\omega^2\partial_{\tau\tau}+(\partial_{xx}+\gamma\partial_{xxxx})^2.  \nonumber
\]
Since the preconditioner must be positive definite, it is sensible
to add a positive constant to the above approximation and hence
choose $\M$ as above. Notice that this $\M$ is self-adjoint (as
required). In addition, its inversion is very simple by using the
Fourier transform. The frequency $\omega$ in this preconditioner is
given through the quasi-Rayleigh quotient in the equation below
(\ref{e:normalKS}).

We first look for time-periodic solutions in the KS equation
(\ref{e:KS2}) with $\gamma=0.054$. At this $\gamma$ value, the KS
equation admits an unstable time-periodic solution
\cite{Papageorgiou1991} . After many random trials (the first
strategy described in the last paragraph of section 2), we arrive at
a successful initial condition
\[  \label{e:fig1ic}
u_0(x, \tau)=-7\sin 3x-3 (\sin 4x-\sin 5x)\sin \tau -\sin x \cos \tau,  \quad 0\le x, \tau \le 2\pi.
\]
In our Newton-CG iterations, we use 64 evenly-spaced grid points
along each of the $x$ and $\tau$ directions. Due to the periodic
conditions of $u(x, \tau)$, we use the discrete (fast) Fourier
transform to evaluate all spatial and temporal derivatives, which
gives spectral accuracy for this algorithm
\cite{Trefethen2000,Boyd2001}. Due to this spectral accuracy, we
find that 64 grid points are already sufficient to yield solutions
accurate within $10^{-10}$. The MATLAB code of this algorithm is
provided in Appendix A (this code is also posted at the author's
homepage: \verb|www.cems.uvm.edu/~jxyang/codes.htm|).

The numerical result of this MATLAB code is displayed in Fig. 1. In
panel (a), the initial condition (\ref{e:fig1ic}) is shown (for two
$\tau$ periods). From this initial condition, the accurate
time-periodic solution obtained by the Newton-CG method is displayed
in panel (b) for two periods of real time $t$, and the accurate
period is found to be $T=1.3297045458$. As one can see from these
two panels, our initial condition differs significantly from the
accurate solution, but the iteration still converges, meaning that
the attraction basin of our Newton-CG method is quite large.
Convergence speed of these Newton-CG iterations is displayed in the
lower panels, where the error versus the number of CG iterations is
plotted in panel (c), while the error versus the time spent is
plotted in panel (d). The error here is defined as
$\mbox{max}|\L_0(u_n)|$, i.e., maximum magnitude of the equation's
residue $\L_0(u_n)$ at the numerical solution $u_n$. Panel (c) shows
that this error drops from the initial value of about 300 to the
final value of $10^{-10}$ under 5000 CG iterations, while panel (d)
shows that this drop of the error from 300 to $10^{-10}$ takes about
6 seconds.

%%%%%%%%%%%%%%%%%%%%%%%%%%%%%%%%%%%%%%%%%%%%%%%%%%%%%%%%%%%%%%%%%%%%%%%%
\begin{figure}[ht!]
\begin{center}
\includegraphics[width=0.75\textwidth]{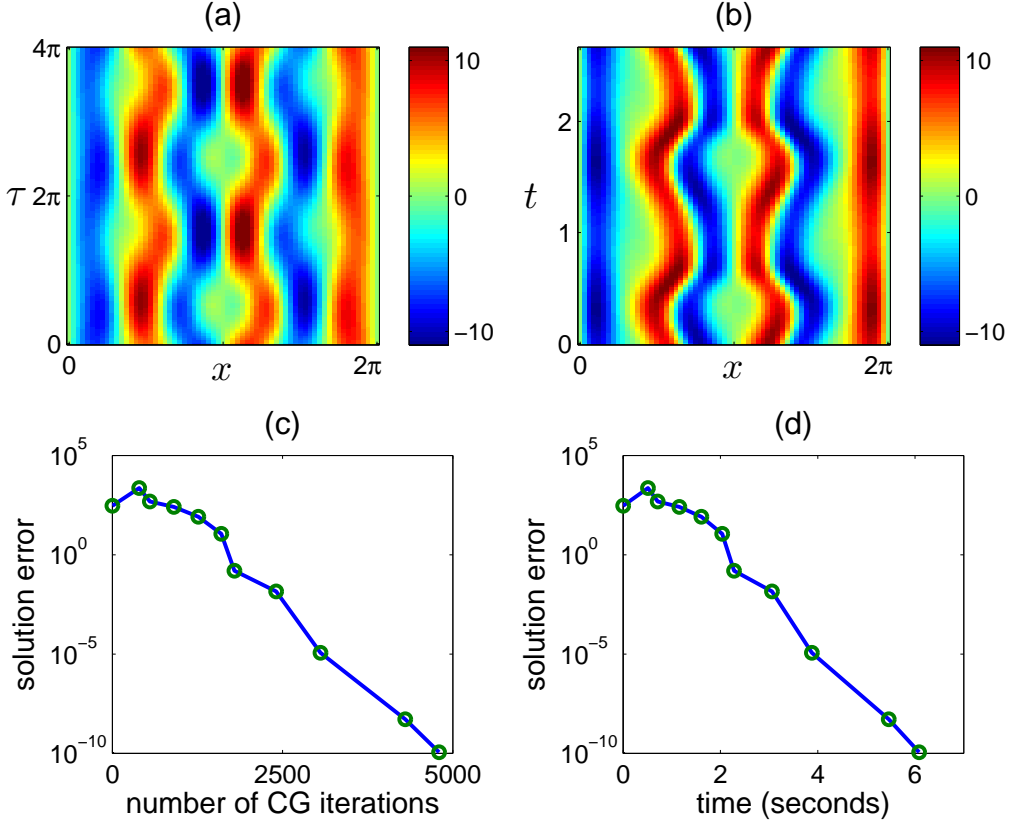} %\kern3em
\end{center}
\caption{Numerical computation of the time-periodic solution in the KS equation (\ref{e:KS2}) at $\gamma=0.054$.
(a) The initial condition $u_0(x, \tau)$ in (\ref{e:fig1ic}). (b) The accurate solution $u(x, t)$. (c) The graph of error versus number of
CG iterations. (d) The graph of error versus the time spent. In (a, b), two time periods are shown. In (c, d), circles are Newton-iteration points.
This figure is produced by the MATLAB code in Appendix A. }
\label{fig1}
\end{figure}
%%%%%%%%%%%%%%%%%%%%%%%%%%%%%%%%%%%%%%%%%%%%%%%%%%%%%%%%%%%%%%%%%%%%%%%%

As the $\gamma$ value decreases, unstable time-periodic solutions
with more complex spatiotemporal structures appear, and
determination of such solutions is supposed to be more challenging
\cite{Cvitanovic2004}. But we find that the Newton-CG method can
handle such solutions with ease as well. To demonstrate, we now take
$\gamma=0.015$. Regarding the initial condition for Newton-CG
iterations, the strategy of random trials has difficulty now due to
the complex structure of the solution. Thus we switch to the
``looking for approximate recurrence" strategy (the second strategy
described in the last paragraph of section 2). Specifically, we
simulate the evolution of the KS equation (\ref{e:KS2}) from the
initial condition $u(x, 0)=-\sin x$. We notice that the evolution
solution in the time interval of $3.49\le t \le 4.21$ is
approximately time-periodic, thus we use this time-segment of the
evolution solution as the initial condition for Newton-CG
iterations. This initial condition proves to converge to an exact
time-periodic solution under Newton-CG iterations, and the numerical
results are displayed in figure 2 (here we use 128 grid points
rather than 64 points along each of the $x$ and $\tau$ directions
since the spatiotemporal structure of the present solution is more
complex). The MATLAB code for this figure is the same as that in
Appendix A, except for the $\gamma$ value, the initial condition
$u_0$, the number of grid points in $(x, \tau)$, and one of the
plotting commands.

Panel (a) of figure 2 shows the difference $u_0(x, t)-u(x,t)$
between our initial condition $u_0(x, t)$ and the accurate
time-periodic solution $u(x, t)$ (for two time periods). One can see
that this difference is not small, meaning that our initial
condition is not very close to the exact solution; but Newton-CG
iterations still converge. The converged (accurate) solution is
displayed in panel (b), and the accurate temporal period is found to
be $T=0.7294854797$. Notice that this time-periodic solution is more
complex than the one in figure 1. Convergence rates of Newton-CG
iterations are shown in panels (c, d), where the error versus the
number of CG iterations and versus time are plotted respectively.
One can see that this error drops from the original 280 to the final
$10^{-9}$ in about 50,000 CG iterations, or 2.6 minutes.

%%%%%%%%%%%%%%%%%%%%%%%%%%%%%%%%%%%%%%%%%%%%%%%%%%%%%%%%%%%%%%%%%%%%%%%%
\begin{figure}[ht!]
\begin{center}
\includegraphics[width=0.75\textwidth]{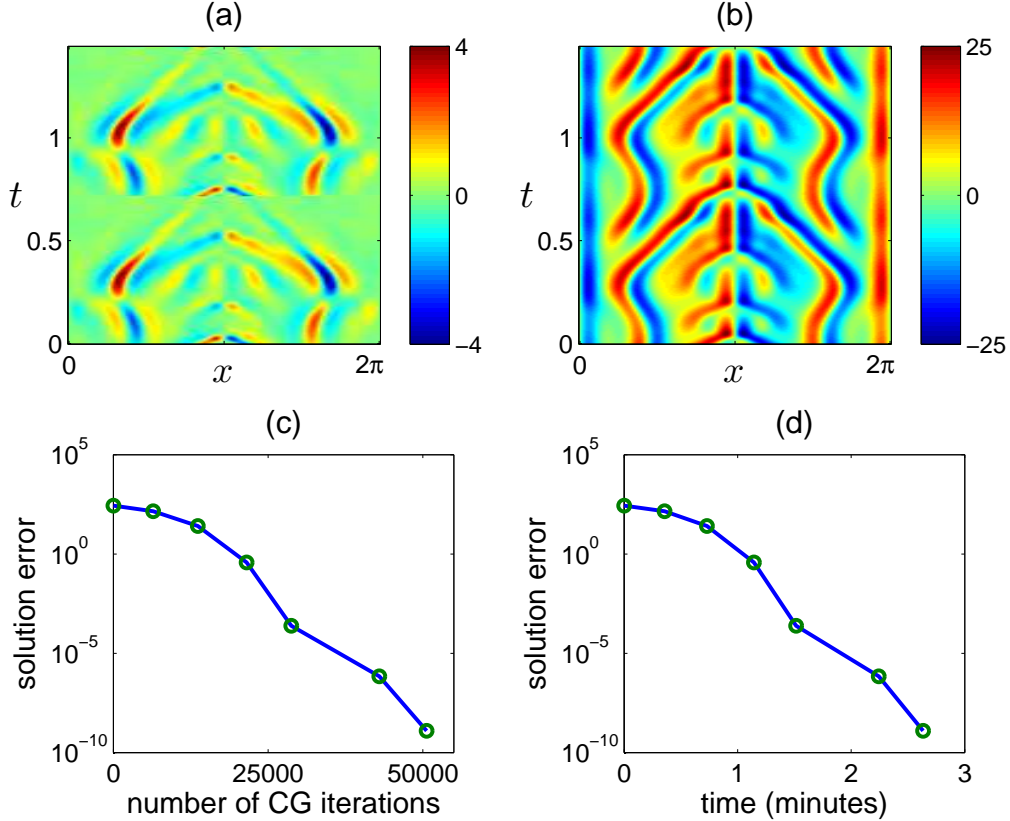}
\end{center}
\caption{Numerical computation of the time-periodic solution in the KS equation (\ref{e:KS2}) at $\gamma=0.015$.
(a) Difference between the initial condition $u_0(x, t)$ and the accurate solution $u(x, t)$.
(b) The accurate solution $u(x, t)$. (c) Error versus the number of
CG iterations. (d) Error versus time. In (a, b), two time periods are shown. In (c, d), circles are Newton-iteration points. }
\label{fig2}
\end{figure}
%%%%%%%%%%%%%%%%%%%%%%%%%%%%%%%%%%%%%%%%%%%%%%%%%%%%%%%%%%%%%%%%%%%%%%%%

\vspace{0.1cm} \textbf{Example 2} \; Our second example is the
cubic-quintic Ginzburg-Landau (CQGL) equation (\ref{e:CQGL}), i.e.,
\[  \label{e:CQGL2}
A_t=\chi A+\gamma A_{xx}-\beta |A|^2A-\delta |A|^4A,
\]
where coefficients $\gamma, \beta, \delta$ are complex, and $\chi$
real (note that if $\chi$ is complex, its imaginary part can be
eliminated by a trivial gauge transformation). This equation admits
relative time-periodic solutions $A(x, t)=e^{i\mu t}U(x, t)$, where
$U(x, t)$ is time-periodic and spatially localized
\cite{Deissler1994,Akhmediev2000,Lopez2005}. Both the propagation
constant $\mu$ and the temporal period $T$ in these solutions are
unknown priori and need to be determined along with the solution
$U(x, t)$. Below we use the numerical algorithm of section 3 to
determine these relative time-periodic solutions.

The CQGL equation (\ref{e:CQGL2}) is of the form (\ref{e:AGL}), thus
the numerical algorithm of section 3 directly applies. In this
algorithm, the function $G$, as defined in (\ref{def:G}), is
\[
G=\chi U+\gamma U_{xx}-\beta |U|^2U-\delta |U|^4U.   \nonumber
\]
Splitting the real and imaginary parts of the complex constants
$\gamma, \beta$, $\delta$ and complex functions $U$, $G$ as
\[
\gamma=\gamma_1+i\gamma_2, \quad \beta=\beta_1+i\beta_2, \quad \delta=\delta_1+i\delta_2,  \quad
U=u+iv, \quad G=g+ih,
\nonumber
\]
we get
\begin{eqnarray*}
g=\gamma_1u_{xx}-\gamma_2v_{xx}+\chi u-(\beta_1u-\beta_2v)(u^2+v^2)-(\delta_1u-\delta_2v)(u^2+v^2)^2, \\
h=\gamma_1v_{xx}+\gamma_2u_{xx}+\chi v-(\beta_1v+\beta_2u)(u^2+v^2)-(\delta_1v+\delta_2u)(u^2+v^2)^2.
\end{eqnarray*}
The linearization operator $\G_1$ of functions $[g,
h]^{\mbox{\scriptsize T}}$ is
\[  \label{e:G1example}
\G_1=\left[ \begin{array}{cc} \gamma_1\partial_{xx}+G_{11} &  -\gamma_2\partial_{xx}+G_{12} \\
\gamma_2\partial_{xx}+G_{21} & \gamma_1\partial_{xx}+G_{22} \end{array}\right],
\]
where
\begin{eqnarray*}
G_{11}&=& \chi-\beta_1(u^2+v^2)-2u(\beta_1u-\beta_2v)-\delta_1(u^2+v^2)^2-4u(\delta_1u-\delta_2v)(u^2+v^2), \\
G_{12}&=& \hspace{0.65cm} \beta_2(u^2+v^2)-2v(\beta_1u-\beta_2v)+\delta_2(u^2+v^2)^2-4v(\delta_1 u-\delta_2v)(u^2+v^2), \\
G_{21}&=& \hspace{0.38cm} -\beta_2(u^2+v^2)-2u(\beta_1v+\beta_2u)-\delta_2(u^2+v^2)^2-4u(\delta_1v+\delta_2u)(u^2+v^2), \\
G_{22}&=& \chi-\beta_1(u^2+v^2)-2v(\beta_1v+\beta_2u)-\delta_1(u^2+v^2)^2-4v(\delta_1v+\delta_2u)(u^2+v^2).
\end{eqnarray*}
The adjoint operator of $\G_1$ is then
$\G_1^A=\G_1^{\mbox{\scriptsize T}}$, i.e.,
\[ \label{e:G1Aexample}
\G_1^A=\left[ \begin{array}{cc} \gamma_1\partial_{xx}+G_{11} & \gamma_2\partial_{xx}+G_{21} \\
-\gamma_2\partial_{xx}+G_{12} & \gamma_1\partial_{xx}+G_{22}
\end{array}\right].
\]
Using the above formulae, the normal equation for Newton corrections
$\Delta \u_n=[\Delta u_n, \Delta v_n]^{\mbox{\scriptsize T}}$ is
\[
\L_{1n}^A \L_{1n} \Delta \u_n = -\L_{1n}^A \L_0(\u_n),
\]
where $\L_1$ is given by equations (\ref{e:L1formula}),
(\ref{e:defP}), (\ref{e:G1example}), $\L_1^A$ given by equations
(\ref{e:L1Adef}), (\ref{e:PA2}), (\ref{e:G1Aexample}), and
$\L_0(\u)$ given by equation (\ref{e:L0u1u2}). This normal equation
will be solved by preconditioned CG iterations.

Regarding the choice of the preconditioner $\M$, we follow the
general guideline in the end of section 2. Specifically, by
retaining only the highest $(x, \tau)$-derivatives of $\Psi$ in the
normal-equation's linear operator $\L_{1}^A \L_{1}\Psi$, we get
\[
\L_{1}^A \L_{1} \approx \left(|\gamma|^2\partial_{xxxx}-\omega^2\partial_{\tau\tau}\right) \textbf{I}_2,  \nonumber
\]
where $\textbf{I}_2$ is a $2\times 2$ identity matrix. Since the
preconditioner must be positive-definite, we then choose the
preconditioner as
\[
\M=\left(c+|\gamma|^2\partial_{xxxx}-\omega^2\partial_{\tau\tau}\right) \textbf{I}_2,
\]
where $c$ is a positive constant (which we set as $c=8$). In
execution, the $\omega$ value in this preconditioner will be
obtained from the numerical solution $\u_n$ through the
quasi-Rayleigh quotient (\ref{e:muomega}).

We now apply the above Newton-CG method to compute relative
time-periodic solutions. First, we choose the parameter values in
the CQGL equation (\ref{e:CQGL2}) as
\[  \label{e:paraGL}
\gamma=0.9-1.1i, \quad  \beta=-3-i, \quad \delta=2.75-i, \quad \chi=-0.1.
\]
For this set of parameter values, the CQGL equation admits a stable
relative-time-periodic and spatially-localized solution
\cite{Deissler1994}. Since this solution is stable, we can use the
time-evolution method (the second strategy in the end of section 2)
to prepare our initial condition for Newton-CG iterations.
Specifically, we numerically simulate the evolution of equation
(\ref{e:CQGL2}) from a Gaussian initial condition $A(x,
0)=e^{-x^2/10}$. This evolution gradually converges to a
relative-time-periodic solution with temporal period of
approximately $7.98$ and propagation constant of approximately $
2.04$. Thus we take the time-segment $200\le t\le 207.98$ of this
solution $A(x, t)$, multiplied by the phase factor of $e^{-2.04it}$,
as our initial condition $U(x,t)$ for Newton-CG iterations. The
$x$-interval is taken as $-50\le x\le 50$, discretized evenly by 512
grid points, and the $\tau$ direction is discretized evenly by 32
grid points. Since the solution is spatially localized and
temporally periodic, we will use discrete Fourier transform to
compute all derivatives. The MATLAB code for this computation is
displayed in Appendix B. This code, together with the initial
condition $U_0(x, t)$, is also posted at the author's homepage.

The numerical result from this MATLAB code is given in figure 3.
This code converges to a time-periodic solution $U(x, t)$, whose
amplitude and phase fields are shown in panels (a, b) (for two
temporal periods). The accurate temporal period is found to be $T=
7.9820986731$, and the accurate propagation constant is
$\mu=2.0422917024$. Convergence speeds of these Newton-CG iterations
are displayed in panels (c, d), where the error versus number of CG
iterations and versus time are plotted. The error here is also
defined as the maximum magnitude of the equation's residue, i.e.,
$\mbox{max}|\L_0(\u_n)|$. Panel (c) shows that this error drops from
the initial value of about 0.3 to the final value below $10^{-10}$
under 2800 CG iterations, while panel (d) shows that this drop of
the error takes about 1.3 minutes.

%%%%%%%%%%%%%%%%%%%%%%%%%%%%%%%%%%%%%%%%%%%%%%%%%%%%%%%%%%%%%%%%%%%%%%%%
\begin{figure}[ht!]
\begin{center}
\includegraphics[width=0.75\textwidth]{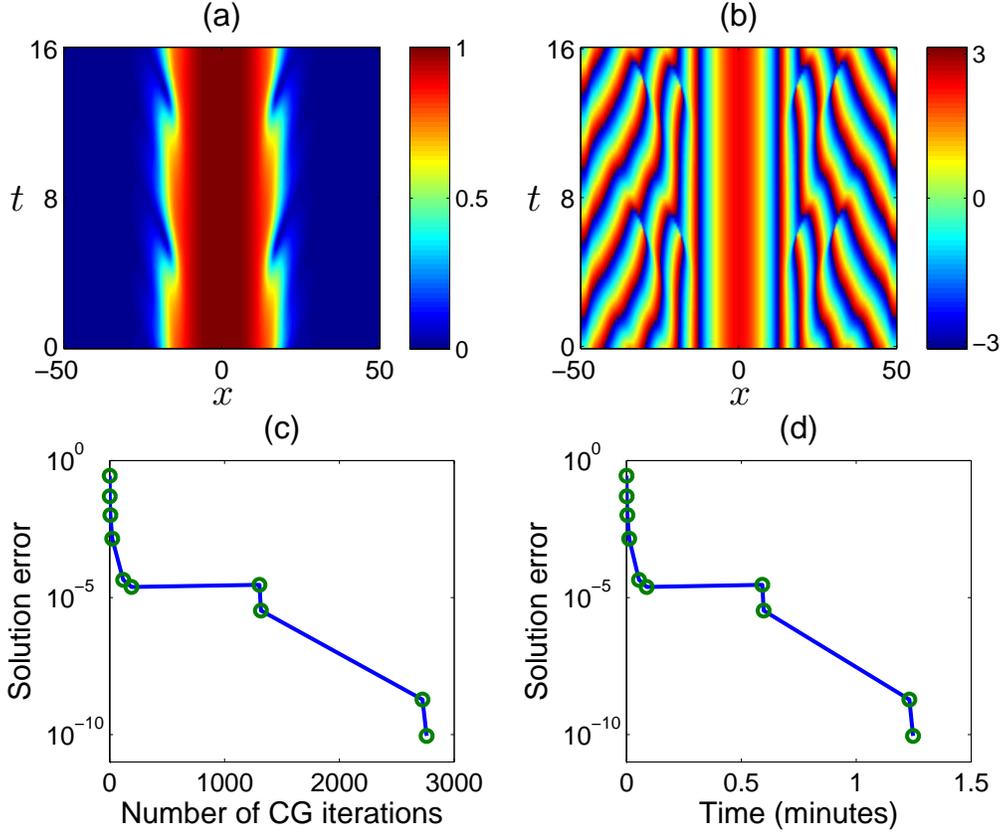}
\end{center}
\caption{Numerical computation of the relative time-periodic solution in the cubic-quintic Ginzburg-Landau equation (\ref{e:CQGL2}) with parameter values
(\ref{e:paraGL}).
(a, b) Amplitude and phase of solution $U(x, t)$.
(c) Error versus the number of CG iterations.
(d) Error versus time. In (a, b), two time periods are shown. In (c, d), circles are Newton-iteration points.
This figure is produced by the MATLAB code in Appendix B.} \label{fig3}
\end{figure}
%%%%%%%%%%%%%%%%%%%%%%%%%%%%%%%%%%%%%%%%%%%%%%%%%%%%%%%%%%%%%%%%%%%%%%%%

When parameters in the CQGL equation (\ref{e:CQGL2}) change, this
stable relative-time-periodic solution in figure 3 can lose its
stability. For instance, when Re($\gamma$) decreases below 0.88 and
the other parameters fixed, this solution would become unstable
\cite{Deissler1994}. Such unstable solutions can be computed
accurately by our Newton-CG method as well. Indeed, starting from
the stable solution of figure 3 and using the continuation method
(the third strategy in the end of section 2), we can track the
entire branch of this solution family parameterized by Re($\gamma$),
and the results are shown in figure 4. Here dependences of the
propagation constant $\mu$ and temporal period $T$ on Re($\gamma$)
are displayed in panels (a, b), and the accurate unstable solution
at Re($\gamma$)$=0.85$ (with error less than $10^{-10}$) is plotted
in panel (c). As can be seen, this continuation by Newton-CG methods
is very suitable for studying bifurcations of time-periodic
solutions.

%%%%%%%%%%%%%%%%%%%%%%%%%%%%%%%%%%%%%%%%%%%%%%%%%%%%%%%%%%%%%%%%%%%%%%%%
\begin{figure}[ht!]
\begin{center}
\includegraphics[width=0.75\textwidth]{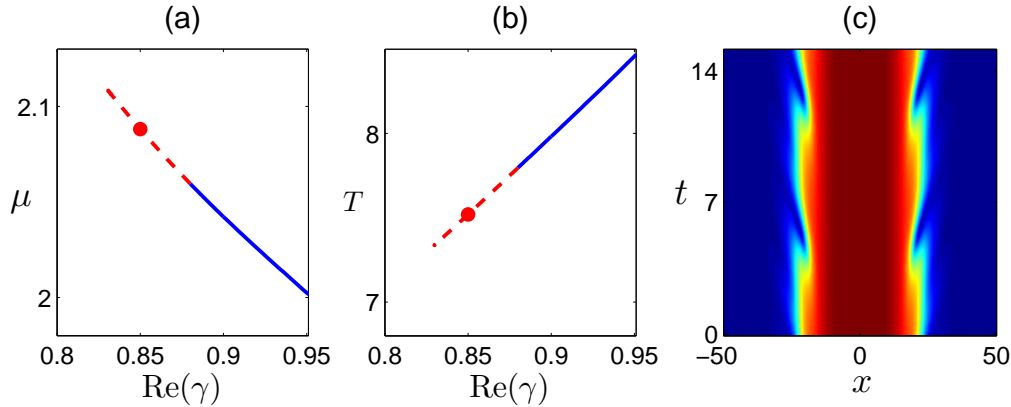}
\end{center}
\caption{Continuation of Newton-CG methods for tracing the entire family of a relative time-periodic solution
in the CQGL equation (\ref{e:CQGL2}), starting from the stable solution in figure 3 (at Re($\gamma$)=0.9).
Here parameters are as given in equation (\ref{e:paraGL}), except
that Re($\gamma$) is allowed to vary. (a, b) Graphs of the
propagation constant $\mu$ and temporal period $T$ versus
Re($\gamma$); (c) Amplitude field of the solution $U(x, t)$ at
Re($\gamma$)$=0.85$ [marked by a red dot in (a, b)]. } \label{fig4}
\end{figure}
%%%%%%%%%%%%%%%%%%%%%%%%%%%%%%%%%%%%%%%%%%%%%%%%%%%%%%%%%%%%%%%%%%%%%%%%

Our numerical algorithms for time-periodic solutions were intended
for dissipative wave equations, such as the KS equation
(\ref{e:KS2}) and the CQGL equation (\ref{e:CQGL2}). But they
certainly apply to systems of ODEs as well. For systems of ODEs, a
number of numerical methods have already been developed to compute
their periodic orbits (see
\cite{Cvitanovic2004,Shooting,Ascher1979,LingWu1987,Gucken2000,AUTO}
for instance). Here we apply our numerical methods to systems of
ODEs and demonstrate their easy computation of periodic orbits in
such systems.

\vspace{0.1cm} \textbf{Example 3} \; The example of systems of ODEs
we consider is the familiar Lorenz equations \cite{Lorenz1963}
\begin{eqnarray}
\frac{dx}{dt} & = & \sigma (y-x), \\
\frac{dy}{dt} & = & rx-y-xz, \\
\frac{dz}{dt} & = & xy-bz,
\end{eqnarray}
where $(x, y, z)$ are real variables of time, and $\sigma, r, b$ are
real constants. These equations admit many types of periodic orbits
in wide ranges of parameter values (see \cite{Strogatz1994} and the
references therein). Below we formulate our numerical algorithm to
compute these periodic orbits.

Periodic orbits in Lorenz equations contain a single unknown
parameter, which is their period. Thus the algorithm of section 2
applies. In this case, the function $\F$ in the algorithm of section
2 is
\[
\F(\u)=\left(\begin{array}{l} \sigma (y-x) \\
rx-y-xz \\
xy-bz \end{array}\right),   \nonumber
\]
where $\u=[x, y, z]^{\mbox{\scriptsize T}}$. The linearization
operator of this function (i.e., the Jacobian) is
\[
\F_1=\left(\begin{array}{ccc} -\sigma & \sigma & 0  \\
r-z & -1 & -x \\
y& x & -b \end{array}\right),   \nonumber
\]
and its adjoint operator is $\F_1^A=\F_1^{\mbox{\scriptsize T}}$.
The quasi-normal Newton-correction equation (\ref{e:normal}) for the
Lorenz equations then is
\[  \label{e:normalLorenz}
\P_n^A \L_{1n} \Delta u_n = -\P_n^A \L_0(u_n),
\]
where
\[
\L_0(\u)=\omega \u_\tau-\F, \quad \L_1\Psi\equiv
\P\Psi-\frac{\langle \u_\tau, \, \P\Psi \rangle} {\langle \u_\tau, \u_\tau \rangle}\u_\tau, \quad \P=\omega \partial_\tau-\F_1, \quad
\P^A=-\omega \partial_\tau-\F_1^A, \quad \omega=\frac{\langle \u_\tau, \F \rangle} {\langle \u_\tau, \u_\tau \rangle},  \nonumber
\]
and we solve it using preconditioned CG iterations.

Regarding the preconditioner, by retaining only the derivative terms
of $\Psi$ in $\P^A \L_{1}\Psi$, we get $\P^A \L_{1} \approx
-\omega^2\partial_{\tau\tau}\textbf{I}_3$, where $\textbf{I}_3$ is
the $3\times 3$ identity matrix. Thus we choose the preconditioner
as
\[
\M=\left(c-\omega^2\partial_{\tau\tau}\right)\textbf{I}_3,
\]
where $c$ is a positive number (which we take as $c=30$).

Now we apply the above Newton-CG method to compute periodic orbits
in the Lorenz equations. As an example, we take $\sigma=10$ and
$b=\frac{8}{3}$, the same values Lorenz used in his pioneering paper
\cite{Lorenz1963}. At these $\sigma$ and $b$ values, a subcritical
Hopf bifurcation occurs at $r=r_H\approx 24.74$, where the pair of
fixed points $(x_c, y_c, z_c)=(\pm \sqrt{b(r-1)}, \pm \sqrt{b(r-1)},
r-1)$ lose their stability when $r>r_H$. When $r<r_H$, an unstable
limit cycle appears \cite{Strogatz1994}. This behavior is
illustrated in figure 5 [panel (a)].

We now compute this unstable limit cycle below $r_H$, with $r=24$
for definiteness. The initial condition for Newton-CG iterations is
chosen by random trials, which yield many successful choices, one of
which being
\[
x_0(\tau)=x_c-2.5\cos(\tau+0.5), \quad y_0(\tau)=y_c+3\sin(\tau-0.4); \quad z_0 (\tau)=z_c-4\cos(\tau-0.3).
\]
We also discretize time evenly by 256 points. The MATLAB code for
this computation is provided in Appendix C.

The numerical outcome of this MATLAB code is given in figure 5
[panels (b,c,d)]. In panel (b), the accurate limit cycle is
displayed. The accurate period is found to be $T=0.6793367642$.
Convergence speeds of Newton-CG iterations are shown in panels (c,
d). We see that the error (defined by $\mbox{max}|\L_0(\u_n)|$ as
before) drops from the initial value of about 12 to the final value
below $10^{-11}$ in 150 CG iterations, or under 0.04 seconds.

%%%%%%%%%%%%%%%%%%%%%%%%%%%%%%%%%%%%%%%%%%%%%%%%%%%%%%%%%%%%%%%%%%%%%%%%
\begin{figure}[ht!]
\begin{center}
\includegraphics[width=0.75\textwidth]{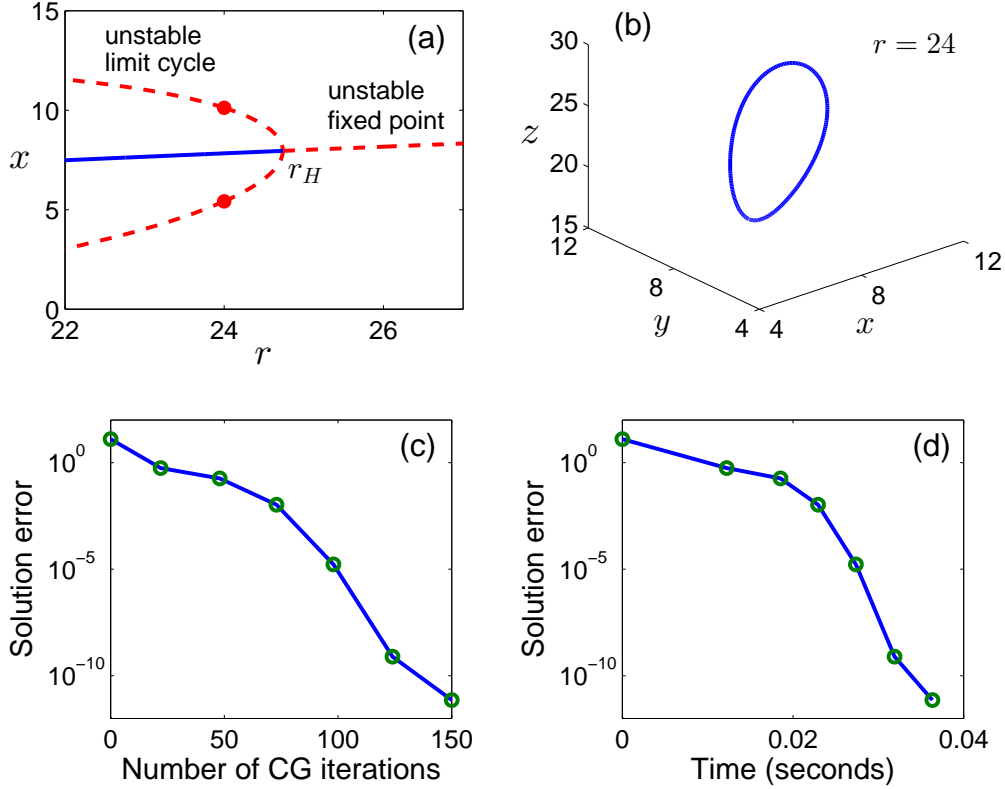}
\end{center}
\caption{Numerical computation of an unstable limit cycle in the Lorenz equations for parameter values of $\sigma=10$, $b=\frac{8}{3}$, and $r=24$.
(a) Bifurcation diagram near the subcritical Hopf bifurcation point $r=r_H\approx 24.74$. (b) Numerically obtained limit cycle.
(c) Error versus number of CG iterations.
(d) Error versus time. This figure (b-d) is produced by the MATLAB code in Appendix C.  } \label{fig5}
\end{figure}
%%%%%%%%%%%%%%%%%%%%%%%%%%%%%%%%%%%%%%%%%%%%%%%%%%%%%%%%%%%%%%%%%%%%%%%%

When $r=28$ (above the Hopf bifurcation point $r_H$), a strange
attractor appears \cite{Lorenz1963}. On this strange attractor, an
infinite number of unstable periodic orbits exist. To look for these
periodic orbits, we apply the above numerical algorithm, starting
from initial conditions
\begin{eqnarray*}
x_0(\tau)=x_c+\sum_{k=1}^2 \left[A_{1k}\cos(k\tau)+B_{1k}\sin(k\tau)\right], \\
y_0(\tau)=y_c+\sum_{k=1}^2 \left[A_{2k}\cos(k\tau)+B_{2k}\sin(k\tau)\right], \\
z_0(\tau)=z_c+\sum_{k=1}^2 \left[A_{3k}\cos(k\tau)+B_{3k}\sin(k\tau)\right],
\end{eqnarray*}
where coefficients $A_{ij}$ and $B_{ij}$ are taken randomly from the
interval $[-10, 10]$. Repeatedly running the MATLAB code of Appendix
C, with $r$ changed to 28, the initial condition changed to the
above random functions, and \verb|errorCG| changed to $10^{-2}$, we
found 20 distinct periodic orbits with period below 10 and accuracy
$10^{-10}$ in 5 minutes.

\section{Summary}
A numerical method was proposed for computing time-periodic and
relative time-periodic solutions in general dissipative wave
systems. Since the temporal period and possibly other additional
internal parameters in the solution are unknown priori, our idea was
to first express those unknown parameters in terms of the solution
through quasi-Rayleigh quotients, so that the resulting
integro-differential equation is for the time-periodic solution
only. Then this integro-differential equation is computed in the
combined spatiotemporal domain by Newton-conjugate-gradient
iterations, where the Newton-correction equation is solved by
preconditioned conjugate gradient iterations. Linearization
operators and their adjoints in the Newton-correction equation were
derived analytically for general systems, so that conjugate gradient
iterations for Newton corrections can be readily implemented.

As numerical examples, we applied this method to the
Kuramoto-Sivashinsky equation and the cubic-quintic Ginzburg-Landau
equation, whose time-periodic or relative time-periodic solutions
with spatially-periodic or spatially-localized profiles were
computed. We also used this method to compute periodic orbits in the
Lorenz equations, since this method applies to systems of ordinary
differential equations as a special case.

Numerical examples showed that first, both stable and unstable
time-periodic solutions can be obtained by this method. Second, the
numerical accuracy of this method is spectral, since we used
spectral differentiation (the discrete Fourier transform) to compute
spatial and temporal derivatives. Thirdly, this method only took
from a fraction of a second to a couple of minutes (on a personal
computer) to find solutions of varying spatiotemporal complexities
to the accuracy of $10^{-10}$, thus this method is fast-converging
and time-efficient. Fourthly, the coding of this method is short and
simple. To make it evident, stand-alone MATLAB codes for our
numerical examples are provided in the appendices.

This proposed method can be a powerful tool for numerically studying
time-periodic (and relative time-periodic) solutions and their
bifurcations in physical systems.

\section*{Acknowledgment}
This work was supported in part by the Air Force Office of
Scientific Research (grant USAF 9550-12-1-0244) and the National
Science Foundation (grant DMS-1311730).

\begin{appendix}

\section{MATLAB code for the Kuramoto-Sivashinsky equation}

In this appendix, we provide the MATLAB code for computing an
unstable time- and space-periodic solution in the
Kuramoto-Sivashinsky equation (\ref{e:KS2}) with $\gamma=0.054$. The
output of this code is shown in figure~1.

\vspace{0.5cm}
\begin{verbatim}
% Newton-CG method for computing time-space-periodic solutions
% in the KS equation: u_t+uu_x+u_{xx}+gamma*u_{xxxx}=0.
% In this code, z represents scaled time tau in the paper.

gamma=0.054; Nx=64; Nz=64; Lx=2*pi; Lz=2*pi; errormax=1e-9; errorCG=1e-4;
dx=Lx/Nx; x=0:dx:Lx-dx;  kx=[0:Nx/2-1  -Nx/2:-1]*2*pi/Lx;
dz=Lz/Nz; z=0:dz:Lz-dz;  kz=[0:Nz/2-1  -Nz/2:-1]*2*pi/Lz;
[X,Z]=meshgrid(x,z); [KX,KZ]=meshgrid(kx,kz); KX2=-KX.*KX+gamma*KX.^4;
u0=-7*sin(3*X)-3*sin(Z).*(sin(4*X)-sin(5*X))-cos(Z).*sin(X); u=u0; % i.c.

tic; nnt=0; ncg=0;    % nnt: # of Newton steps;  ncg: # of CG iterations
while 1               % Newton-CG iterations for periodic solutions start
  nnt=nnt+1;
  ufft=fft2(u);
  F=-real(u.*ifft2(i*KX.*ufft)+ifft2(KX2.*ufft));
  uz=real(ifft2(i*KZ.*ufft));
  omega=sum(sum(uz.*F))/sum(sum(uz.*uz));
  L0u=omega*uz-F;
  uerror(nnt)=max(max(abs(L0u))); uerror(nnt)
  numcg(nnt)=ncg; time(nnt)= toc;
  if uerror(nnt) < errormax
    break
  end

  P=@(W)  real(ifft2(( omega*i*KZ+KX2).*fft2(W))+ifft2(i*KX.*fft2(u.*W)));
  PA=@(W) real(ifft2((-omega*i*KZ+KX2).*fft2(W))-u.*ifft2(i*KX.*fft2(W)));

  c=30; fftM=omega^2*KZ.*KZ+KX2.*KX2+c;            % Preconditioner
  du=0*Z;                                          % CG iterations start
  R=-PA(L0u);
  MinvR=real(ifft2(fft2(R)./fftM));
  R2=sum(sum(R.*MinvR)); R20=R2;
  D=MinvR;
  while (R2 > R20*errorCG^2)
     PD=P(D);
     L1D=PD-sum(sum(uz.*PD))/sum(sum(uz.*uz))*uz;
     PAL1D=PA(L1D);
     a=R2/sum(sum(D.*PAL1D));
     du=du+a*D;
     R=R-a*PAL1D;
     MinvR=real(ifft2(fft2(R)./fftM));
     R2old=R2;
     R2=sum(sum(R.*MinvR));
     b=R2/R2old;
     D=MinvR+b*D;
     ncg=ncg+1;
  end                                              % CG iterations end
  u=u+du;
end                                           % Newton-CG iterations end

% plotting of numerical results
subplot(221); imagesc(x, [z z+Lz], [u0; u0]); axis xy; colorbar;
xlabel('x'); ylabel('\tau','rotation',0); title('(a)');
subplot(222); imagesc(x, [z z+Lz]/omega, [u; u]); axis xy; colorbar;
xlabel('x'); ylabel('t','rotation',0); title('(b)');
subplot(223); semilogy(numcg, uerror, numcg, uerror, 'o');
xlabel('number of CG iterations'); ylabel('solution error'); title('(c)');
subplot(224); semilogy(time, uerror, time, uerror, 'o');
xlabel('time (seconds)'); ylabel('solution error'); title('(d)');
format long; period=2*pi/omega
\end{verbatim}

%%%%%%%%%%%%%%%%%%%%%%%%%%%%%%%%%%%%%%%%%%%%%%%%%%%%%%%%%%%%%%%%%%%%%%%%%%%%%%%%%%%%%%%%%%%%%%%%%%%%%%%%%%%%%%%%%%%%%%

\section{MATLAB code for the cubic-quintic Ginzburg-Landau equation}

In this appendix, we provide the MATLAB code for computing a
(stable) relative-time-periodic and space-localized solution in the
cubic-quintic Ginzburg-Landau equation (\ref{e:CQGL2}) with
parameters (\ref{e:paraGL}). The initial condition
\verb|U0_fig3.mat| in this code is obtained from simulating the CQGL
equation from a Gaussian initial condition $A(x, 0)=e^{-x^2/10}$
(see text for details). The MATLAB data for this initial condition
can be found at the author's homepage
\verb|www.cems.uvm.edu/~jxyang/codes.htm|. From this initial
approximation (whose error is about 0.3), the following MATLAB code
then drives the error below $10^{-10}$, and the output of this code
is shown in figure~3. Note that during MATLAB implementation of the
algorithm, real functions $u$ and $v$ are recombined into $U=u+iv$,
so that they can be computed simultaneously for numerical efficiency
and compact coding. Because of it, real operators $\P, \P^A, \L_1$
and $\L_1^A$ in the algorithm are adjusted into complex operators,
and some inner products are expressed through these complex
functions.

\vspace{0.5cm}
\begin{verbatim}
% Newton-CG method for computing time-periodic and space-localized solutions
% in the CQGL equation: At-gamma*Axx+beta*|A|^2A+delta*|A|^4A-chi*A=0.
% In this code, z represents scaled time tau, and A=U*exp(i*mu*t).

load U0_fig3.mat;   % this data contains initial condition U(x, z)
Lx=100; Nx=512; Lz=2*pi; Nz=32; errormax=1e-10; errorCG=1e-4;
dx=Lx/Nx; x=-Lx/2:dx:Lx/2-dx;  kx=[0:Nx/2-1  -Nx/2:-1]*2*pi/Lx;
dz=Lz/Nz; z=0:dz:Lz-dz;        kz=[0:Nz/2-1  -Nz/2:-1]*2*pi/Lz;
[X,Z]=meshgrid(x,z);           [KX,KZ]=meshgrid(kx,kz);   KX2=KX.*KX;

gamma=0.9-1.1i; beta=-3-i; delta=2.75-i; chi=-0.1;
gamma1=real(gamma); gamma2=imag(gamma); beta1=real(beta); beta2=imag(beta);
delta1=real(delta); delta2=imag(delta);

tic; nnt=0; ncg=0;    % nnt: # of Newton steps;  ncg: # of CG iterations
while 1               % Newton-CG iterations for periodic solutions start
  nnt=nnt+1;
  u=real(U); v=imag(U); U2=u.*u+v.*v; U4=U2.*U2;
  G=gamma*ifft2(-KX2.*fft2(U))-(beta*U2+delta*U4-chi).*U;
  Ut=ifft2(i*KZ.*fft2(U)); ut=real(Ut); vt=imag(Ut);
  produv=2*sum(sum(u.*v)); produtvt=2*sum(sum(ut.*vt));
  mu   = sum(sum(v.*imag(G)-u.*real(G)))/produv;
  omega= sum(sum(ut.*imag(G)+vt.*real(G)))/produtvt;
  L0U=omega*Ut+i*mu*U-G;
  Uerror(nnt)=max(max(abs(L0U))); Uerror(nnt)
  numcg(nnt)=ncg; time(nnt)= toc;
  if Uerror(nnt) < errormax
      break
  end

  betaU1=beta1*u-beta2*v;    betaU2=beta1*v+beta2*u;
  deltaU1=delta1*u-delta2*v; deltaU2=delta1*v+delta2*u;
  G11=chi-beta1*U2-betaU1*2.*u-delta1*U4-deltaU1*4.*u.*U2;
  G12=   +beta2*U2-betaU1*2.*v+delta2*U4-deltaU1*4.*v.*U2;
  G21=   -beta2*U2-betaU2*2.*u-delta2*U4-deltaU2*4.*u.*U2;
  G22=chi-beta1*U2-betaU2*2.*v-delta1*U4-deltaU2*4.*v.*U2;
  Dxx=@(F)   ifft2(-KX2.*fft2(F));
  Dtxx=@(F)  ifft2(( omega*i*KZ+gamma1*KX2).*fft2(F));
  DtxxA=@(F) ifft2((-omega*i*KZ+gamma1*KX2).*fft2(F));

  P=@(F)  Dtxx(real(F))-G11.*real(F)-(mu+G12).*imag(F)+gamma2*Dxx(imag(F)) ...
     +i*( (mu-G21).*real(F)-gamma2*Dxx(real(F))+Dtxx(imag(F))-G22.*imag(F) );
  PA=@(F) DtxxA(real(F))-G11.*real(F)+(mu-G21).*imag(F)-gamma2*Dxx(imag(F)) ...
     +i*( -(mu+G12).*real(F)+gamma2*Dxx(real(F))+DtxxA(imag(F))-G22.*imag(F) );

  L1= @(F) P(F)-sum(sum(imag(Ut.*P(F))))/produtvt*Ut ...
              +sum(sum(real(U.*P(F))))/produv*i*U;
  L1A=@(F) PA(F)-sum(sum(real(conj(F).*Ut)))/produtvt*PA(vt+i*ut) ...
              -sum(sum(imag(conj(F).*U)))/produv*PA(u-i*v);

  c=8; fftM=omega^2*KZ.*KZ+abs(gamma)^2*KX2.*KX2+c;     % Preconditioner
  dU=0*Z;                          % CG iterations start
  R=-L1A(L0U);
  MinvR=ifft2(fft2(R)./fftM);
  R2=sum(sum(real(conj(R).*MinvR))); R20=R2;
  D=MinvR;
  while (R2 > R20*errorCG^2)
      L2D=L1A(L1(D));
      a=R2/sum(sum(real(conj(D).*L2D)));
      dU=dU+a*D;
      R=R-a*L2D;
      MinvR=ifft2(fft2(R)./fftM);
      R2old=R2;
      R2=sum(sum(real(conj(R).*MinvR)));
      b=R2/R2old;
      D=MinvR+b*D;
      ncg=ncg+1;
  end                              % CG iterations end
  U=U+dU;
end                                  % Newton-CG iterations end

% plotting of numerical results
subplot(221); imagesc(x, [z z+Lz]/omega, abs([U;U])); axis xy; colorbar;
xlabel('x'); ylabel('t'); title('(a)');
subplot(222); imagesc(x, [z z+Lz]/omega, angle([U;U])); axis xy; colorbar;
xlabel('x'); ylabel('t'); title('(b)');
subplot(223); semilogy(numcg, Uerror, numcg, Uerror, 'o');
xlabel('number of CG iterations'); ylabel('solution error'); title('(c)');
subplot(224); semilogy(time/60, Uerror, time/60, Uerror, 'o');
xlabel('time (minutes)'); ylabel('solution error'); title('(d)');
format long; period=2*pi/omega
mu
\end{verbatim}

%%%%%%%%%%%%%%%%%%%%%%%%%%%%%%%%%%%%%%%%%%%%%%%%%%%%%%%%%%%%%%%%%%%%%%%%%%%%%%%%%%%%%%%%%%%%%%%%%%%%%%%%%%%%%%%%%%%%%%

\section{MATLAB code for the Lorenz equation}

In this appendix, we provide the MATLAB code for computing an
unstable limit cycle in the Lorenz equations with $\sigma=10, b=8/3$
and $r=24$. This limit cycle is located below the subcritical Hopf
bifurcation point $r_H\approx 24.74$. The output of this code is
shown in figure~5(b-d).

\vspace{0.5cm}

\begin{verbatim}
% Newton-CG method for computing limit cycles in the Lorenz equations.

L=2*pi; N=256; errormax=1e-10; errorCG=1e-4;
dtau=L/N; tau=(0:dtau:L-dtau)';
ktau=[0:N/2-1 -N/2:-1]'*2*pi/L; Ktau=[ktau ktau ktau];

sigma=10; b=8/3; r=24; xc=sqrt(b*(r-1)); yc=xc; zc=r-1;
x=xc-2.5*cos(tau+0.5); y=yc+3*sin(tau-0.4); z=zc-4*cos(tau-0.3); u=[x y z];

tic; nnt=0; ncg=0;     % nnt: # of Newton steps;  ncg: # of CG iterations
while 1                % Newton-CG iterations for limit cycles start
    nnt=nnt+1;
    F=[sigma*(y-x)  r*x-y-x.*z  x.*y-b*z];
    utau=real(ifft(i*Ktau.*fft(u)));
    omega=sum(sum(utau.*F))/sum(sum(utau.*utau));
    L0u=omega*utau-F;
    uerror(nnt)=max(max(abs(L0u))); uerror(nnt)
    numcg(nnt)=ncg; time(nnt)= toc;
    if uerror(nnt) < errormax
        break
    end

    P=@(W)  omega*real(ifft(i*Ktau.*fft(W)))   ...
           -[-sigma*W(:,1)+sigma*W(:,2),       ...
             (r-z).*W(:,1)-W(:,2)-x.*W(:,3),   ...
              y.*W(:,1)+x.*W(:,2)-b*W(:,3)];

    PA=@(W) -omega*real(ifft(i*Ktau.*fft(W)))        ...
           -[-sigma*W(:,1)+(r-z).*W(:,2)+y.*W(:,3),  ...
              sigma*W(:,1)-W(:,2)+x.*W(:,3),         ...
              -x.*W(:,2)-b*W(:,3)];

    c=30; fftM=omega^2*Ktau.*Ktau+c;      % Preconditioner
    du=0*u;                               % CG iterations start
    R=-PA(L0u);
    MinvR=real(ifft(fft(R)./fftM));
    R2=sum(sum(R.*MinvR)); R20=R2;
    D=MinvR;
    while (R2 > R20*errorCG^2)
        PD=P(D);
        L1D=PD-sum(sum(utau.*PD))/sum(sum(utau.*utau))*utau;
        PAL1D=PA(L1D);
        a=R2/sum(sum(D.*PAL1D));
        du=du+a*D;
        R=R-a*PAL1D;
        MinvR=real(ifft(fft(R)./fftM));
        R2old=R2;
        R2=sum(sum(R.*MinvR));
        beta=R2/R2old;
        D=MinvR+beta*D;
        ncg=ncg+1;
    end                                   % CG iterations end
    u=u+du;
    x=u(:,1); y=u(:,2); z=u(:,3);
end                               % Newton-CG iterations end

% plotting of numerical results
subplot(222); plot3(x, y, z); xlabel('x'); ylabel('y'); zlabel('z');
title('(b)'); axis([4 12 4 12 15 30]); view([-40 30])
subplot(223); semilogy(numcg, uerror, numcg, uerror, 'o');
xlabel('number of CG iterations'); ylabel('solution error'); title('(c)');
subplot(224); semilogy(time, uerror, time, uerror, 'o');
xlabel('time (seconds)'); ylabel('solution error'); title('(d)');
format long; period=2*pi/omega
\end{verbatim}

\end{appendix}

\end{document}